\documentclass[onecolumn,letterpaper,aps,prd,longbibliography,superscriptaddress,showpacs,floatfix]{revtex4-1}

\usepackage{graphicx}	
\usepackage{amssymb}
\usepackage{amsmath}
\usepackage{resizegather}
\usepackage{tabularx}
\usepackage{xspace}	

\newcommand{\gevc}{\mbox{GeV/$c$}~}

\newcommand{\jpsi}{\mbox{$J/\psi~$}}

\newcommand{\pp}{\mbox{$p$+$p$~}}
\newcommand{\cc}{\mbox{$c\bar{c}~$}}
\newcommand{\bb}{\mbox{$b\bar{b}~$}}
\newcommand{\full}{\mbox{$\sqrt{s}=$ 510 GeV~}}

\newcommand{\dcar}{\mbox{$\textrm{DCA}_{\rm R}$}~}

\newcommand{\bfrac}{\mbox{$F_{B{\rightarrow}J/\psi}~$}}

\newcommand{\rarr}{\mbox{$\rightarrow$~}}

\begin{document}

\title{Measurements of $B \rightarrow J/\psi$ at forward rapidity in 
$p$$+$$p$ collisions at $\sqrt{s}=510$ GeV}

\newcommand{\abilene}{Abilene Christian University, Abilene, Texas 79699, USA}
\newcommand{\augie}{Department of Physics, Augustana University, Sioux Falls, South Dakota 57197, USA}
\newcommand{\banaras}{Department of Physics, Banaras Hindu University, Varanasi 221005, India}
\newcommand{\barc}{Bhabha Atomic Research Centre, Bombay 400 085, India}
\newcommand{\baruch}{Baruch College, City University of New York, New York, New York, 10010 USA}
\newcommand{\bnlcoll}{Collider-Accelerator Department, Brookhaven National Laboratory, Upton, New York 11973-5000, USA}
\newcommand{\bnlphys}{Physics Department, Brookhaven National Laboratory, Upton, New York 11973-5000, USA}
\newcommand{\caucr}{University of California-Riverside, Riverside, California 92521, USA}
\newcommand{\charlesczech}{Charles University, Ovocn\'{y} trh 5, Praha 1, 116 36, Prague, Czech Republic}
\newcommand{\chonbuk}{Chonbuk National University, Jeonju, 561-756, Korea}
\newcommand{\ciae}{Science and Technology on Nuclear Data Laboratory, China Institute of Atomic Energy, Beijing 102413, People's Republic of China}
\newcommand{\cns}{Center for Nuclear Study, Graduate School of Science, University of Tokyo, 7-3-1 Hongo, Bunkyo, Tokyo 113-0033, Japan}
\newcommand{\colorado}{University of Colorado, Boulder, Colorado 80309, USA}
\newcommand{\columbia}{Columbia University, New York, New York 10027 and Nevis Laboratories, Irvington, New York 10533, USA}
\newcommand{\czechtech}{Czech Technical University, Zikova 4, 166 36 Prague 6, Czech Republic}
\newcommand{\debrecen}{Debrecen University, H-4010 Debrecen, Egyetem t{\'e}r 1, Hungary}
\newcommand{\elte}{ELTE, E{\"o}tv{\"o}s Lor{\'a}nd University, H-1117 Budapest, P{\'a}zm{\'a}ny P.~s.~1/A, Hungary}
\newcommand{\ewha}{Ewha Womans University, Seoul 120-750, Korea}
\newcommand{\fsu}{Florida State University, Tallahassee, Florida 32306, USA}
\newcommand{\gsu}{Georgia State University, Atlanta, Georgia 30303, USA}
\newcommand{\hanyang}{Hanyang University, Seoul 133-792, Korea}
\newcommand{\hiroshima}{Hiroshima University, Kagamiyama, Higashi-Hiroshima 739-8526, Japan}
\newcommand{\howard}{Department of Physics and Astronomy, Howard University, Washington, DC 20059, USA}
\newcommand{\ihepprot}{IHEP Protvino, State Research Center of Russian Federation, Institute for High Energy Physics, Protvino, 142281, Russia}
\newcommand{\illuiuc}{University of Illinois at Urbana-Champaign, Urbana, Illinois 61801, USA}
\newcommand{\inrras}{Institute for Nuclear Research of the Russian Academy of Sciences, prospekt 60-letiya Oktyabrya 7a, Moscow 117312, Russia}
\newcommand{\instpasczech}{Institute of Physics, Academy of Sciences of the Czech Republic, Na Slovance 2, 182 21 Prague 8, Czech Republic}
\newcommand{\isu}{Iowa State University, Ames, Iowa 50011, USA}
\newcommand{\jaea}{Advanced Science Research Center, Japan Atomic Energy Agency, 2-4 Shirakata Shirane, Tokai-mura, Naka-gun, Ibaraki-ken 319-1195, Japan}
\newcommand{\jyvaskyla}{Helsinki Institute of Physics and University of Jyv{\"a}skyl{\"a}, P.O.Box 35, FI-40014 Jyv{\"a}skyl{\"a}, Finland}
\newcommand{\karoly}{K\'aroly R\'oberts University College, H-3200 Gy\"ngy\"os, M\'atrai \'ut 36, Hungary}
\newcommand{\kek}{KEK, High Energy Accelerator Research Organization, Tsukuba, Ibaraki 305-0801, Japan}
\newcommand{\korea}{Korea University, Seoul, 136-701, Korea}
\newcommand{\kurchatov}{National Research Center ``Kurchatov Institute", Moscow, 123098 Russia}
\newcommand{\kyoto}{Kyoto University, Kyoto 606-8502, Japan}
\newcommand{\labllr}{Laboratoire Leprince-Ringuet, Ecole Polytechnique, CNRS-IN2P3, Route de Saclay, F-91128, Palaiseau, France}
\newcommand{\lahorelums}{Physics Department, Lahore University of Management Sciences, Lahore 54792, Pakistan}
\newcommand{\lawllnl}{Lawrence Livermore National Laboratory, Livermore, California 94550, USA}
\newcommand{\losalamos}{Los Alamos National Laboratory, Los Alamos, New Mexico 87545, USA}
\newcommand{\lund}{Department of Physics, Lund University, Box 118, SE-221 00 Lund, Sweden}
\newcommand{\maryland}{University of Maryland, College Park, Maryland 20742, USA}
\newcommand{\mass}{Department of Physics, University of Massachusetts, Amherst, Massachusetts 01003-9337, USA}
\newcommand{\michigan}{Department of Physics, University of Michigan, Ann Arbor, Michigan 48109-1040, USA}
\newcommand{\muhlenberg}{Muhlenberg College, Allentown, Pennsylvania 18104-5586, USA}
\newcommand{\myongji}{Myongji University, Yongin, Kyonggido 449-728, Korea}
\newcommand{\nagasaki}{Nagasaki Institute of Applied Science, Nagasaki-shi, Nagasaki 851-0193, Japan}
\newcommand{\nara}{Nara Women's University, Kita-uoya Nishi-machi Nara 630-8506, Japan}
\newcommand{\natmephi}{National Research Nuclear University, MEPhI, Moscow Engineering Physics Institute, Moscow, 115409, Russia}
\newcommand{\newmex}{University of New Mexico, Albuquerque, New Mexico 87131, USA}
\newcommand{\nmsu}{New Mexico State University, Las Cruces, New Mexico 88003, USA}
\newcommand{\ohio}{Department of Physics and Astronomy, Ohio University, Athens, Ohio 45701, USA}
\newcommand{\ornl}{Oak Ridge National Laboratory, Oak Ridge, Tennessee 37831, USA}
\newcommand{\orsay}{IPN-Orsay, Univ.~Paris-Sud, CNRS/IN2P3, Universit\'e Paris-Saclay, BP1, F-91406, Orsay, France}
\newcommand{\peking}{Peking University, Beijing 100871, People's Republic of China}
\newcommand{\pnpi}{PNPI, Petersburg Nuclear Physics Institute, Gatchina, Leningrad region, 188300, Russia}
\newcommand{\riken}{RIKEN Nishina Center for Accelerator-Based Science, Wako, Saitama 351-0198, Japan}
\newcommand{\rikjrbrc}{RIKEN BNL Research Center, Brookhaven National Laboratory, Upton, New York 11973-5000, USA}
\newcommand{\rikkyo}{Physics Department, Rikkyo University, 3-34-1 Nishi-Ikebukuro, Toshima, Tokyo 171-8501, Japan}
\newcommand{\saispbstu}{Saint Petersburg State Polytechnic University, St.~Petersburg, 195251 Russia}
\newcommand{\seoulnat}{Department of Physics and Astronomy, Seoul National University, Seoul 151-742, Korea}
\newcommand{\stonybrkc}{Chemistry Department, Stony Brook University, SUNY, Stony Brook, New York 11794-3400, USA}
\newcommand{\stonycrkp}{Department of Physics and Astronomy, Stony Brook University, SUNY, Stony Brook, New York 11794-3800, USA}
\newcommand{\sungskku}{Sungkyunkwan University, Suwon, 440-746, Korea}
\newcommand{\tenn}{University of Tennessee, Knoxville, Tennessee 37996, USA}
\newcommand{\titech}{Department of Physics, Tokyo Institute of Technology, Oh-okayama, Meguro, Tokyo 152-8551, Japan}
\newcommand{\tsukuba}{Center for Integrated Research in Fundamental Science and Engineering, University of Tsukuba, Tsukuba, Ibaraki 305, Japan}
\newcommand{\vandy}{Vanderbilt University, Nashville, Tennessee 37235, USA}
\newcommand{\weizmann}{Weizmann Institute, Rehovot 76100, Israel}
\newcommand{\wigner}{Institute for Particle and Nuclear Physics, Wigner Research Centre for Physics, Hungarian Academy of Sciences (Wigner RCP, RMKI) H-1525 Budapest 114, POBox 49, Budapest, Hungary}
\newcommand{\yonsei}{Yonsei University, IPAP, Seoul 120-749, Korea}
\newcommand{\zagreb}{University of Zagreb, Faculty of Science, Department of Physics, Bijeni\v{c}ka 32, HR-10002 Zagreb, Croatia}
\affiliation{\abilene}
\affiliation{\augie}
\affiliation{\banaras}
\affiliation{\barc}
\affiliation{\baruch}
\affiliation{\bnlcoll}
\affiliation{\bnlphys}
\affiliation{\caucr}
\affiliation{\charlesczech}
\affiliation{\chonbuk}
\affiliation{\ciae}
\affiliation{\cns}
\affiliation{\colorado}
\affiliation{\columbia}
\affiliation{\czechtech}
\affiliation{\debrecen}
\affiliation{\elte}
\affiliation{\ewha}
\affiliation{\fsu}
\affiliation{\gsu}
\affiliation{\hanyang}
\affiliation{\hiroshima}
\affiliation{\howard}
\affiliation{\ihepprot}
\affiliation{\illuiuc}
\affiliation{\inrras}
\affiliation{\instpasczech}
\affiliation{\isu}
\affiliation{\jaea}
\affiliation{\jyvaskyla}
\affiliation{\karoly}
\affiliation{\kek}
\affiliation{\korea}
\affiliation{\kurchatov}
\affiliation{\kyoto}
\affiliation{\labllr}
\affiliation{\lahorelums}
\affiliation{\lawllnl}
\affiliation{\losalamos}
\affiliation{\lund}
\affiliation{\maryland}
\affiliation{\mass}
\affiliation{\michigan}
\affiliation{\muhlenberg}
\affiliation{\myongji}
\affiliation{\nagasaki}
\affiliation{\nara}
\affiliation{\natmephi}
\affiliation{\newmex}
\affiliation{\nmsu}
\affiliation{\ohio}
\affiliation{\ornl}
\affiliation{\orsay}
\affiliation{\peking}
\affiliation{\pnpi}
\affiliation{\riken}
\affiliation{\rikjrbrc}
\affiliation{\rikkyo}
\affiliation{\saispbstu}
\affiliation{\seoulnat}
\affiliation{\stonybrkc}
\affiliation{\stonycrkp}
\affiliation{\sungskku}
\affiliation{\tenn}
\affiliation{\titech}
\affiliation{\tsukuba}
\affiliation{\vandy}
\affiliation{\weizmann}
\affiliation{\wigner}
\affiliation{\yonsei}
\affiliation{\zagreb}
\author{C.~Aidala} \affiliation{\losalamos} \affiliation{\michigan} 
\author{N.N.~Ajitanand} \affiliation{\stonybrkc} 
\author{Y.~Akiba} \email[PHENIX Spokesperson: ]{akiba@rcf.rhic.bnl.gov} \affiliation{\riken} \affiliation{\rikjrbrc} 
\author{R.~Akimoto} \affiliation{\cns} 
\author{J.~Alexander} \affiliation{\stonybrkc} 
\author{M.~Alfred} \affiliation{\howard} 
\author{K.~Aoki} \affiliation{\kek} \affiliation{\riken} 
\author{N.~Apadula} \affiliation{\isu} \affiliation{\stonycrkp} 
\author{H.~Asano} \affiliation{\kyoto} \affiliation{\riken} 
\author{E.T.~Atomssa} \affiliation{\stonycrkp} 
% ...
\collaboration{PHENIX Collaboration} \noaffiliation

\date{\today}


\begin{abstract}

We report the first measurement of the fraction of $J/\psi$ mesons 
coming from $B$-meson decay ($F_{B{\rightarrow}J/\psi}$) in $p$+$p$ 
collisions at $\sqrt{s}=$ 510 GeV.  The measurement is performed using 
the forward silicon vertex detector and central vertex detector at 
PHENIX, which provide precise tracking and distance-of-closest-approach 
determinations, enabling the statistical separation of $J/\psi$ due to 
$B$-meson decays from prompt $J/\psi$.  The measured value of 
$F_{B{\rightarrow}J/\psi}$ is 8.1\%$\pm$2.3\% (stat)$\pm$1.9\% (syst) 
for $J/\psi$ with transverse momenta $0<p_T<5$ GeV/$c$ and rapidity 
$1.2<|y|<2.2$. The measured fraction $F_{B{\rightarrow}J/\psi}$ at 
PHENIX is compared to values measured by other experiments at higher 
center of mass energies and to fixed-order-next-to-leading-logarithm 
and color-evaporation-model predictions.  The $b\bar{b}$ cross section per 
unit rapidity ($d\sigma/dy(pp{\rightarrow}b\bar{b})$) extracted from the 
obtained $F_{B{\rightarrow}J/\psi}$ and the PHENIX inclusive $J/\psi$ 
cross section measured at 200 GeV scaled with color-evaporation-model 
calculations, at the mean $B$ hadron rapidity $y={\pm}1.7$ in 510 GeV 
$p$$+$$p$ collisions, is $3.63^{+1.92}_{-1.70}\mu$b. It is 
consistent with the fixed-order-next-to-leading-logarithm calculations.

\end{abstract}

	
\maketitle

\section{introduction}

The measurement of bottom ($B$) mesons in $p$+$p$ and $p$+$\bar{p}$ 
collisions is of interest to constrain the total bottom cross section 
as well as test our understanding of bottom quark production mechanisms 
and hadronization. There are extensive direct measurements of various 
$B$ mesons, as well as measurements of $B \rightarrow J/\psi$ 
contributions over a broad range in \jpsi transverse momentum and 
rapidity from the Tevatron in $p$+$\bar{p}$ at $\sqrt{s}$ = 1.8, 1.96 
TeV~\cite{Abe:1993hr, Abachi:1996jq, Acosta:2004yw} and the Large 
Hadron Collider (LHC) in $p$+$p$ at $\sqrt{s}=7$--13 
TeV~\cite{Abelev:2012gx,Khachatryan:2010yr,lhcb8TeV,Aaij:2015rla,Aad2016}. 
In contrast, measurements from UA1 in $p$$+$$\bar{p}$ at 
$\sqrt{s}=630$~GeV~\cite{UA1} are statistically limited and only for 
$p_T(J/\psi) >$ 5 GeV$/c$. Adding new measurements at lower energies 
and covering different kinematic regions is valuable for testing 
perturbative quantum chromodynamics (pQCD) calculations and 
constraining production mechanisms.
 
The Relativistic Heavy Ion Collider (RHIC) provides $p$$+$$p$ collisions 
at $\sqrt{s}$ = 200, 500 and 510 GeV, which extends the kinematic reach 
for bottom measurements. At these smaller energies, bottom production is 
dominated by gluon-gluon fusion, while higher energy bottom production 
contains a larger fraction of flavor excitation and gluon splitting 
processes~\cite{Norrbin:2000zc}. The STAR experiment measured $B 
\rightarrow J/\psi$ at midrapidity for \jpsi $p_{T} > $ 5 GeV/$c$ in 
$p$+$p$ at $\sqrt{s}$ = 200 GeV~\cite{Adamczyk:2012ey}. Our measurement 
at forward rapidity and $p_{T}$ within 0--5 GeV$/c$ in $\sqrt{s}=510$ 
GeV $p$+$p$ collisions at PHENIX can provide the validation of parton 
distribution functions (PDF) in a different gluon fractional momentum 
range $5 \times 10^{-4}<x_{Bj}<1\times10^{-2}$. As the highest center of mass 
energy accessed by RHIC collisions, bottom measurements at 
$\sqrt{s}=510$ GeV will also help us understand the energy dependence 
from RHIC to LHC energies.

Inclusive \jpsi production has a component referred to as ``prompt'', 
that includes direct \jpsi production, as well as decays from $\psi'$ 
and $\chi_{c}$. The term ``prompt'' is in contrast to ``nonprompt'', 
which specifically refers to production through more long-lived decay 
parent hadrons (i.e. $B$ mesons). The nonprompt \jpsi component that 
comes from the decay of $B$ mesons, provides a clean channel to measure 
$B$-meson yields. At forward rapidities, the time dilation of the $B$ 
lifetime leads to a larger displacement from the event vertex before 
decaying to \jpsi. We use this displacement to separate \jpsi 
originating from $B$-meson decay from prompt \jpsi through measurement 
of the decay particle's distance of closest approach (DCA) to the 
primary event vertex.

In this paper, the ratio of \jpsi from $B$-meson decays to inclusive 
\jpsi (\bfrac) is determined for \jpsi kinematics in the range of 
$0<p_{T}<5$ \gevc and rapidity $1.2<|y|<2.2$ through DCA distributions 
in $p$+$p$ collisions at \full, using the PHENIX muon arms plus the 
forward and central silicon vertex tracker detectors. The $b\bar{b}$ 
cross section per unit rapidity at the mean $B$ hadron rapidity $y = 
\pm1.7$ in 510 GeV $p$+$p$ collisions is extracted from the obtained 
\bfrac and the PHENIX inclusive \jpsi cross section measured at 200 GeV, 
scaled with color-evaporation-model ({\sc cem}) calculations 
\cite{PhysRevLett.95.122001}.

The paper is organized as follows. Section~\ref{sec: Experiment setup} 
discusses the PHENIX detector setup for this analysis, in particular 
the central and forward silicon vertex detectors which are used for the 
primary vertex and the DCA determination. 
Section~\ref{sec: Analysis Procedure} describes the data reconstruction 
and simulation setup, signal and background determination, and fitting 
procedure. The acceptance$\times$efficiency correction factor to 
achieve final results and the systematic uncertainty evaluation are 
discussed in Section~\ref{sec: Analysis Procedure} as well. The results 
and interpretation are discussed in Section~\ref{sec: 
results_and_discussions} and the conclusions are summarized in 
Section~\ref{sec: Summary}.

               \section{Experimental setup}
\label{sec: Experiment setup}

\begin{figure}[tbh]
	\includegraphics[width=0.96\linewidth]{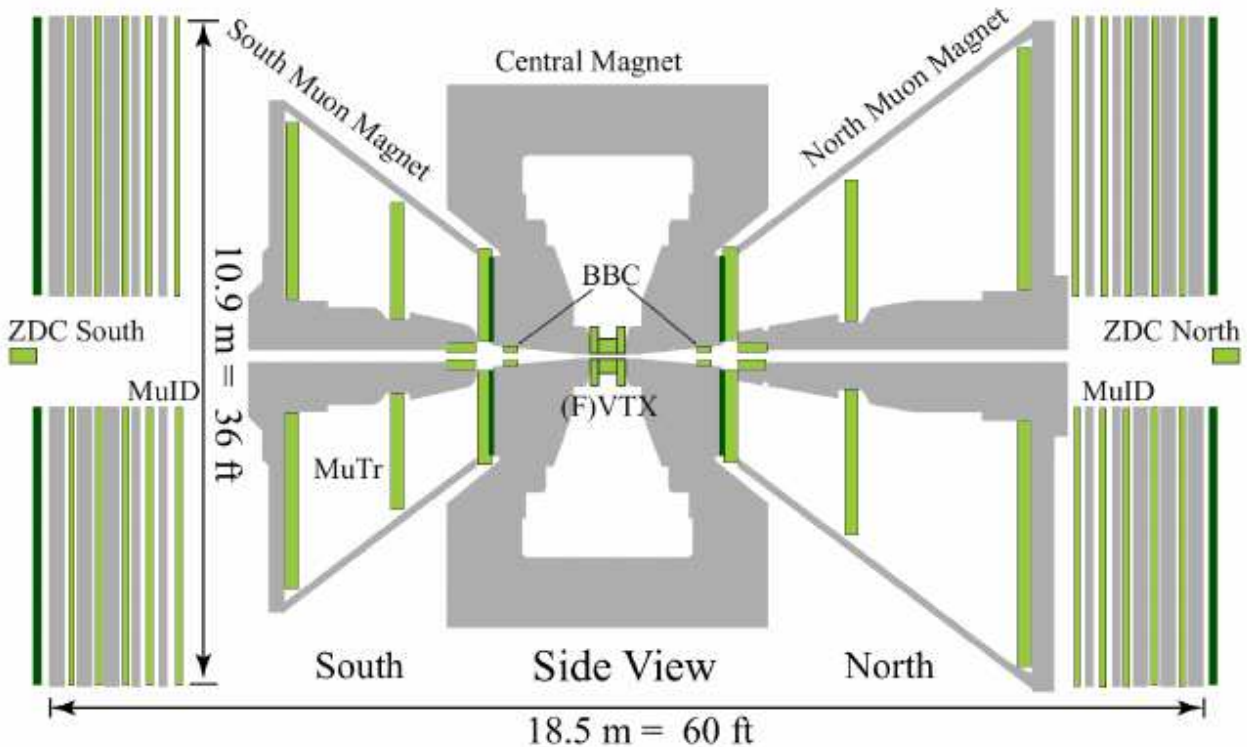} 
\caption{\label{fig:phenix_2012} The PHENIX detector setup for 
the 510 GeV $p$+$p$ data taking in 2012.}
\end{figure}

The data set used in this analysis is from the 2012 run of $p$+$p$ at 
$\sqrt{s}$ = 510 GeV and the detector configuration of PHENIX for that 
running period is shown in Fig.~\ref{fig:phenix_2012}. For this 
measurement, the beam-beam counters (BBC)~\cite{Mallen2003499}, the muon 
arm spectrometers~\cite{Akikawa2003537}, the central silicon vertex 
detector (VTX)~\cite{vtxnim1,vtxnim2} and the forward silicon vertex 
detector (FVTX)~\cite{fvtx_nim} are used. The BBC detector, which 
comprises 128 quartz \v{C}erenkov counters with a pseudorapidity coverage 
of $3.0<|\eta|<3.9$, determines when a collision event has taken place. 
The BBC provides the minimum-bias (MB) trigger, by requiring a  
coincidence between at least one hit in both the positive and negative 
acceptance of the BBC.

The PHENIX muon detectors are divided into the North ($1.2<y<2.4$) and 
the South ($-2.2<y<-1.2$) arms. Each muon arm spectrometer has full 
azimuthal coverage and is composed of hadron absorbers, a muon tracker 
(MuTr) which resides in a radial field magnet, and a muon identifier 
(MuID). The MuTr comprises three cathode strip wire chamber stations 
inside a magnet which provides a radial magnetic field with an 
integrated bending power of around 0.8 T$\cdot$m. The MuTr measures 
track momentum $p$ with a resolution of $\delta p/p \approx$ 0.05 at 
$p<10$ GeV$/c$. The hadron absorber comprises 19 cm of copper, 60 cm of 
iron and 36.2 cm of stainless steel along the beam axis. The absorbers 
are situated in front of the MuTr to provide hadron (mostly pion and 
kaon) rejection. The MuTr has a position resolution at each station of 
around 100 $\mu$m, which, together with a precisely determined vertex, 
results in a mass resolution of around 95 MeV for dimuon pairs within 
the \jpsi mass region and $0<p_{T}(J/\psi)<5$ GeV$/c$. The downstream 
MuID comprises five sandwiched planes of Iarocci proportional tubes and 
steel.  The MuTr+MuID system together with the steel absorbers have 
approximately 10 interaction lengths of material. In this analysis, the 
dimuon trigger is used which requires two muon-like trajectories 
(defined as a ``road") passing through at least three MuID planes with at 
least one reaching the last plane of the MuID.

The VTX (installed in 2011) comprises two inner pixel layers and two 
outer strip layers distributed from 2.5 cm to 14.0 cm along the radial 
direction, covering $\Delta \varphi \approx 5.0$ radians in azimuth and 
$|z({\rm VTX})| < 10$ cm along the $z$ axis (beam direction). The radii 
of the inner silicon pixel detectors are 2.5 and 5.0 cm, and the radii 
of the outer silicon strip detectors are on average 10.0 and 14.0 cm. 
Each pixel of the inner VTX layers covers a 50 $\mu$m $\times$ 450 
$\mu$m active area~\cite{vtxnim1, vtxnim2}. The FVTX, installed in front 
of the hadron absorbers in 2012, comprises 4 silicon disks perpendicular 
to the beam axis and placed at approximately $z$ = $\pm 20.1$ cm, $\pm 
26.1$ cm, $\pm 32.2$ cm and $\pm 38.2$ cm. The rapidity coverage of the 
FVTX overlaps the muon arm coverage. Each FVTX disk comprises 48 
individual silicon sensors (wedges) and each wedge contains two columns 
of strips that each span an azimuthal segmentation of $3.75 ^{o}$. The 
column comprises mini-strips with 75 $\mu$m width in the radial 
direction. The strip length in the azimuthal ($\varphi$) direction 
varies from 3.4 mm at the inner radius to 11.5 mm at the outer radius 
for the largest stations~\cite{fvtx_nim}. Tracks passing through the 
forward muon arms are unlikely to pass through the VTX outer strip 
layers due to the angular acceptance of the strips. In addition, the 
two inner pixel layers can help improve the DCA resolutions as they are 
closer to the vertex and have finer pixel sizes compared to the outer 
strip layers. Therefore, for track reconstruction with the combined 
FVTX+VTX detectors, only the two inner pixel layers in the VTX are 
used.

The FVTX enhances the existing muon arm tracking performance in several 
ways. The FVTX helps reject hadrons that undergo multiple scattering or 
decay inside the hadron absorber by requiring a good joint fit of FVTX 
and MuTr tracks. It also provides a better opening angle determination 
than the MuTr alone can provide, which results in an improved mass 
resolution for dimuon pairs. Finally, the additional precision tracking 
added in front of the hadron absorber by the FVTX makes the measurement 
of displaced tracks possible when combined with a determination of the 
primary vertex position.

\begin{figure*}[tb]
	\includegraphics[width=0.43\linewidth]{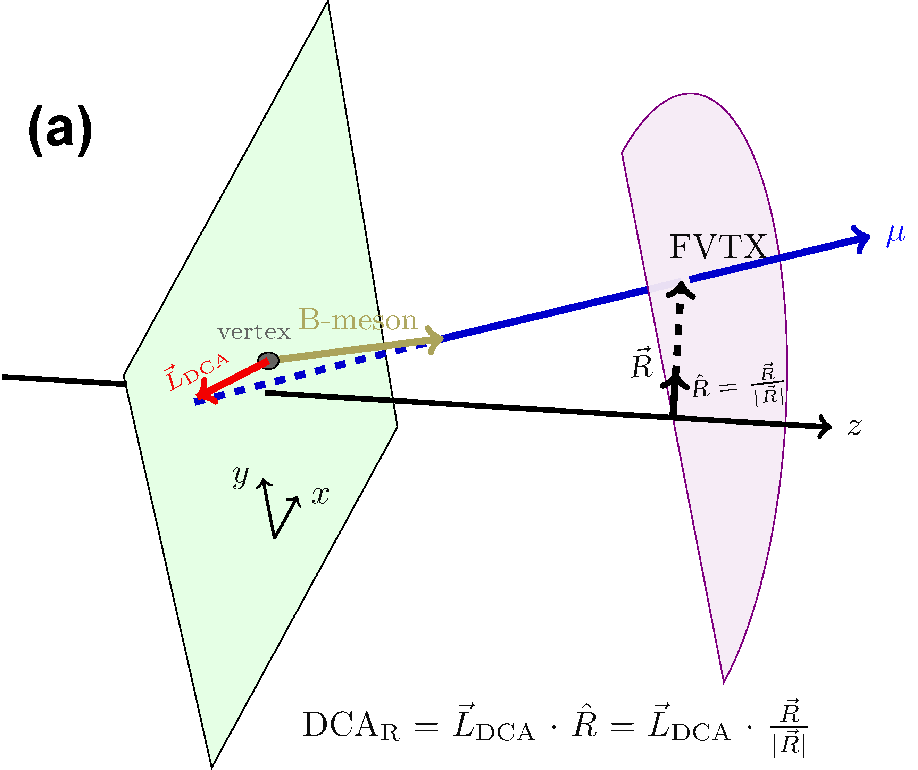}
	\includegraphics[width=0.53\linewidth]{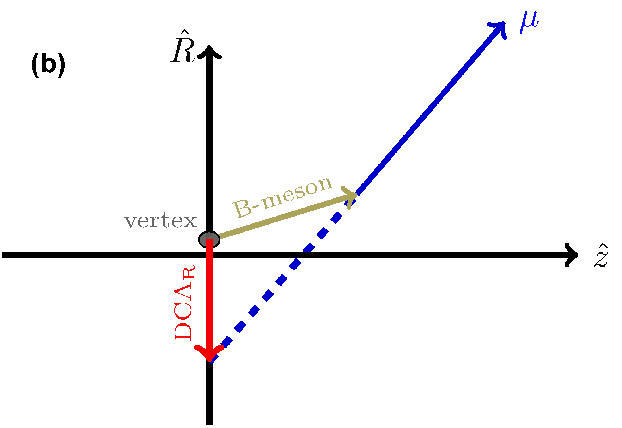}
\caption{\label{fig:define_dca} (a) 3D and (b) projection of a 
muon from a $B$ meson to \jpsi to dimuon decay to the transverse vertex 
plane ($x$-$y$) and definition of \dcar.}
\end{figure*}

Due to limited resolutions in the $z$ and azimuthal $\varphi$ components 
of the FVTX detector, the separation of prompt and decay muons is 
realized with the FVTX using the DCA measurement instead of measuring 
the displaced vertex of decayed muons. Because the FVTX has better 
resolution in the radial direction than in the azimuthal direction, the 
radial DCA (DCA$_{\rm R}$) is the primary variable used in this 
analysis. The primary vertex is reconstructed using all FVTX and VTX 
tracks which pass the track quality cut $\chi^{2}/\rm{NDF} < 4$,
where NDF is the number of degrees of freedom. 
Figure~\ref{fig:define_dca} illustrates the projection of a muon from a 
$B$ meson to \jpsi decay in the transverse vertex plan and how to 
calculate the $\rm{DCA}_{\rm R}$. A track reconstructed in the FVTX is 
extrapolated to the transverse plane ($x$-$y$) at the $z$ location of 
the primary collision vertex. DCA is defined as the vector $\vec{L}_{\rm 
DCA}$ formed between this intersection point and the $x$-$y$ collision 
vertex point in the same transverse plane of the collision vertex. The 
\dcar is the component of the DCA which is measured in the same radial 
direction as the FVTX strips,

\begin{equation}
\dcar \equiv \vec{L}_{\rm DCA} \cdot \hat{R}= \vec{L}_{\rm DCA} \cdot \frac{\vec{R}}{|\vec{R}|}.
\end{equation}
Prompt particles from the primary collision vertex have a symmetric 
\dcar distribution centered at zero, with the width determined by the 
intrinsic detector and vertex resolutions, while the shape is 
asymmetric for decay particles from a displaced decay vertex. As 
illustrated in Fig.~\ref{fig:define_dca}(b), the definition of \dcar 
results in an asymmetric distribution for muons from $B \rightarrow 
J/\psi$ decay due to the projection onto the transverse $x$-$y$ plane 
of the primary vertex.  This is confirmed by the full simulation shown 
in Section~\ref{sec: Simulation Setup}.

\section{Analysis Procedure}
\label{sec: Analysis Procedure}

This analysis starts with the identification of good \jpsi candidates by 
selecting dimuon pairs found by the MuTr that are matched to MuID 
tracks. Separately, track finding is performed in the FVTX/VTX system 
where reconstructed tracks are required to contain at least one FVTX hit 
and a total of at least three FVTX+VTX hits. Then, for each 
reconstructed MuTr track, the FVTX/VTX tracks are searched for potential 
matches.

The collision point is determined from VTX and FVTX tracks. First, 
regions where there is a concentration of track crossings are 
determined. The center of gravity of each of these regions defines a 
collision point. For each region, the center of gravity is used to 
initiate a minimization of the vector sum of the DCAs of the tracks. 
During the minimization, tracks with large displacements are removed to 
improve the fidelity of the final vertex reconstruction. The vertex 
determination in each event is strongly affected by the small VTX and 
FVTX track multiplicities in $p$+$p$ collisions. Events containing \bb 
decay products can also skew the vertex determination. Therefore, in 
this analysis we take advantage of the beam stability in $x$ and $y$ 
during the fill (5--12 hours) and use the measured average $x$ and $y$ 
position of all events in the fill to determine our primary $x$ and $y$ 
vertex. The spread of the primary $x$ and $y$ vertex position based on 
the beam spot size is around 80 $\mu$m in RMS. The $z$ position is 
still determined on an event-by-event basis for events that have a 
VTX+FVTX track multiplicity $\ge$2.  Events with smaller multiplicity 
are thrown out. For events with more than one reconstructed vertex, the 
vertex with the best reconstruction quality is selected as the primary 
vertex. For the reconstructed events, we obtain an average $z$ 
resolution of approximately 180 $\mu$m in 510 GeV $p$+$p$ collisions. 
After matching to the FVTX tracks, the \dcar is determined using the 
MuTr+MuID+FVTX combined track fit and the primary vertex location.

The next step in the analysis is to characterize the \dcar of muons 
from prompt \jpsi decay and \jpsi from $B$-meson decay through 
simulation. The final analysis step uses a fit function for the muon 
\dcar spectra that includes the prompt $J/\psi$, \jpsi from $B$-meson 
decay, and background components to extract the fraction of \jpsi from 
$B$-meson decay in the data, using a log-likelihood fit. Details of the 
analysis procedure are explained step by step in the following 
sections.

\subsection{ Data Quality Assurance}
\label{sec: Data QA}

The precise primary $z$ vertex reconstruction is limited by the VTX 
acceptance and therefore only events within a $z$ vertex ($z_{\rm VTX}$) 
window of (-10, 10) cm are selected for this analysis. Events with 
poorly determined primary $z$ vertices are removed by requiring less 
than 400 $\mu$m calculated uncertainty on the $z$-vertex. Runs without 
an accurately determined average $x$, $y$ position of the beam center 
are rejected.  The number of events with MB and dimuon triggers 
surviving after these vertex selections is 3.5 $\times$ 10$^{9}$, which 
is equivalent to a total integrated luminosity of 0.47 $pb^{-1}$. The 
event rejection fraction is around 67\%.

During the 2012 $p$+$p$ run, there were some areas of the FVTX detector 
which were not yet operational due to various electronics issues. When 
the FVTX-MuTr matching algorithm tries to find an FVTX track in a dead 
area, there is a tendency for it to match to a track in a live region 
neighboring the dead one instead, pulling the matching distributions 
away from the central value of 0. Because of this tendency to pull 
tracks away from a symmetric distribution, fiducial cuts are applied to 
remove tracks that point to the vicinity of a dead region in the FVTX 
detector.

Detector misalignments can shift the projected track position in the 
vertex plane and thus distort the \dcar distributions. Before proceeding 
with the data analysis, alignment corrections are applied to the data in 
two stages, before and after the track reconstruction. The 
pre-production alignment left residual $\varphi$-dependent 
misalignments, which were up to 100 $\mu$m in certain detector regions. 
Tilts which shift the FVTX silicon sensors out of the normal $x$-$y$ 
plane were corrected in a post-production alignment procedure, reducing 
the final misalignment values to less than 30 $\mu$m.

A final verification of the FVTX alignment to the VTX, which is the most 
critical alignment for DCA analyses, is performed using real data. 
Tracks which show MuID activity in the fourth Iarocci tube plane 
(gap), but not in the last 
gap are first selected.  The majority of these tracks are from stopped 
hadrons, which are predominantly prompt particles, and provide a high 
statistics sample for studying alignment. Events with a large vertex 
uncertainty, tracks next to dead areas, and bad quality FVTX tracks are 
removed from this sample. To remove the hadron decay component, a 
minimum longitudinal momentum cut of ($p_{z}>4$ GeV/$c$) is required. 
After the misalignment corrections described above are applied, the 
\dcar is then extracted for these tracks and checked for any indications 
of residual misalignments. The mean of these distributions is found to 
be flat along the $\varphi$ direction (within the measurement precision) 
and the overall offsets of the distributions are within 30 $\mu$m in 
both arms. These offset values are much smaller than the detector 
position resolution. Variations of the \dcar mean and spread which could 
occur if there were beam instability, detector, trigger or 
acceptance$\times$efficiency changes, are checked by examining the \dcar 
distributions as a function of run and BBC instantaneous rate. The mean 
values of the \dcar distributions across all runs are found to be within 
one standard deviation (of the intrinsic \dcar distribution width) after 
quality assurance checks.

\begin{table*}
\caption{ Quality cuts for \jpsi candidates in \pp collisions.}
\begin{ruledtabular} 
\begin{tabular}{ccl}
Variable (Meaning) & $1.2<|y|<2.2$ \\ 
\hline
$|z_{\rm VTX}|$ (collision vertex measured by the FVTX/VTX) & $<10$ cm  \\ 
$|z_{\rm VTX} \ \rm{uncertainty}|$ (collision vertex uncertainty measured by the FVTX/VTX) & $<400$ $\mu$m  \\ 
$p \cdot DG0$ (Track momentum times the spatial difference between  & $<80$ GeV/$c$ $\cdot$ cm  \\
the MuTr track and MuID track at the first MuID layer) &   \\ 
$p \cdot DDG0$ (Track momentum times the slope difference between & $<40$ GeV/$c$ $\cdot$ $^{\circ}$   \\
the MuTr track and MuID track at the first MuID layer) &  \\ 
$\chi^{2}_{\rm MuTr}$ ($\chi^{2}/\rm{NDF}$ of the MuTr track) & $<10$ \\
$\chi^{2}_{\rm MuID}$ ($\chi^{2}/\rm{NDF}$ of the MuID road) & $<3$ \\
Track $\chi^2_{\rm{FVTX-MuTr}}$ ($\chi^{2}/\rm{NDF}$ of the FVTX-MuTr matching $\mu$ track) & $<5$ \\ 
Radial residual between FVTX and MuTr projection at FVTX station 4 & $< 3 \sigma$ \\
Azimuthal residual between FVTX and MuTr projection at FVTX station 4 & $< 3 \sigma$  \\
Last gap (Last MuID plane that the $\mu$ track penetrated) & = 4  \\
nidhits (Number of hits in the MuID, out of the maximum 10 ) & $>6$  \\ 
ntrhits (Number of hits in the MuTr, out of the maximum 16 ) & $>11$ \\ 
nfvtxhits (Number of hits in the FVTX+VTX, out of the maximum 6 ) & $>2$ \\ 
$|p_{z}| (\rm{GeV}/c)$ (Momentum of the $\mu$ along the beam axis) & $>3$ \\ 
dimuon pair vertex $\chi^{2}/\rm{NDF}$ &  $ < 3$ 
\label{tab:quality_cut}
\end{tabular} 
\end{ruledtabular}
\end{table*}

\subsection{ \jpsi Reconstruction}
\label{sec: jpsi Reconstruction}

Tracks formed in the MuTr are required to contain at least 12 (out of 
16) hits in the various cathode strip planes. We start with a loose 
quality cut $\chi^{2}/\rm{NDF}<10$ on the MuTr tracks to make sure all 
potentially good tracks are included in the analysis. The MuTr tracks 
which reach the last gap of the MuID and have longitudinal momentum 
$>3$ GeV/$c$ are treated as muon track candidates. Muon candidates in 
this analysis need to have good associations between the MuTr track and 
the MuID road in both position and angle. The momentum-dependent 
position and angle differences between the MuTr track and the MuID road 
are required to be within three standard deviations as calculated using 
the Kalman Filter track fitting and error propagation method. In 
addition, the associated MuID road should contain at least 6 (out of 
10) hits in different MuID planes. Because the MuID road is not 
included in the fully reconstructed tracks, we apply a tighter quality 
cut which is $\chi^{2}/\rm{NDF}<3$.

Good matching between the FVTX tracks and the MuTr+MuID tracks is also 
required. This requirement helps remove mis-reconstructed and bad 
quality tracks as well as some hadronic background. The matched FVTX 
tracks should contain at least 3 (out of 6) FVTX+VTX hits. The 
differences in azimuthal angle, polar angle and radial distance between 
matched FVTX and MuTr+MuID combined tracks are required to be within 
three standard deviations as determined by the Kalman Filter fits and 
error propagation. Fits on the combined FVTX+MuTr tracks should satisfy 
$\chi^{2}/\rm{NDF}<5$.  Dimuon pairs are created from muons passing all 
the quality cuts. A slightly different selection which requires at 
least one muon of the dimuon pair passing through the quality cuts is 
tested. No bias is found as consistent results are achieved between the 
two selections. The fit of the vertex point plus the two muon tracks 
with opposite charges must satisfy $\chi^{2}/\rm{NDF}<3$ to ensure the 
two muon tracks are not separated by more than 1 mm. The complete set 
of quality cuts is listed in Table~\ref{tab:quality_cut}.

\begin{figure}[!htb]
	\includegraphics[width=1.0\linewidth]{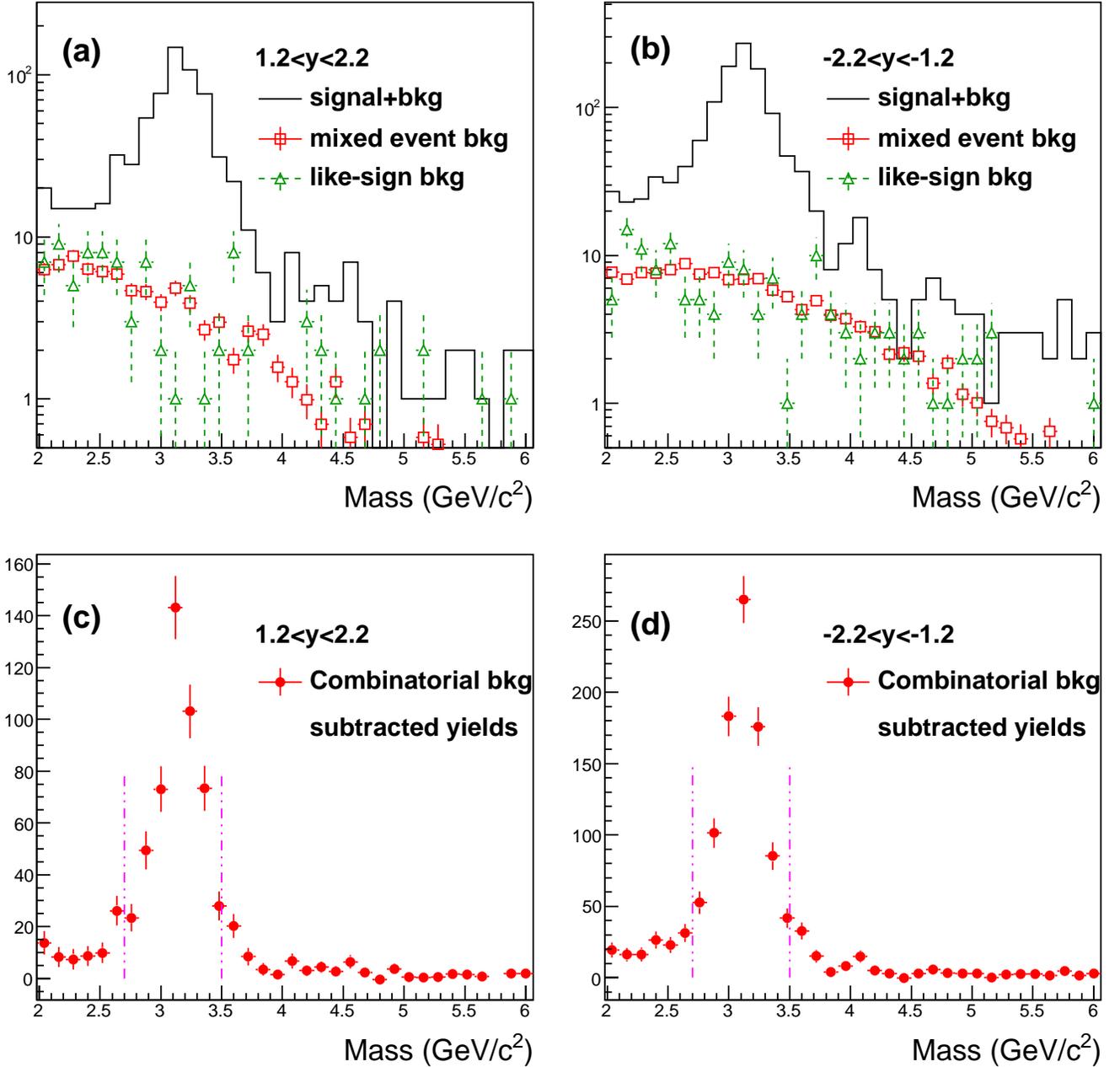}
\caption{\label{fig:jpsi_mass} The invariant mass of dimuons in the 
(a,c) $1.2<y<2.2$ and (b,d) $-2.2<y<-1.2$ regions. Raw yields (black 
solid), the combinatorial background using mixed events (red open 
rectangular) and like-sign dimuon pairs (green open triangle) are shown 
in panels (a) and (b). The combinatorial background subtracted yields are 
shown in panels (c) and (d). The magenta dashed lines represent the mass 
cut used to select \jpsi candidates.}
\end{figure}

Raw yields of the invariant mass of dimuon pairs after applying the 
quality cuts are shown in Figs.~\ref{fig:jpsi_mass}(a) and (b). A 
smaller number of events is measured in the forward than the backward 
rapidity due to larger MuTr dead areas and lower MuID efficiency in the 
forward rapidity region during this data taking period. These spectra 
contain a combination of \jpsi events, combinatorial background (random 
combinations of reconstructed tracks within an event) and heavy flavor 
background. The heavy flavor background determination will be discussed 
in Section~\ref{sec:hf_background}. Two methods are used to extract the 
combinatorial background. One uses the like-sign dimuon pairs within 
events, and the other uses the unlike-sign dimuon pairs in mixed 
events. To match the yields of the analyzed mixed events to the (same) 
events, a normalization scale $\textrm{Norm}_{\rm mix}$, defined in Eq. 
(\ref{eq:norm_mix}), is applied to the mass distribution of dimuon 
pairs and muon \dcar distribution in mixed events:

\begin{equation}
\label{eq:norm_mix}
{\rm Norm}_{\rm mix} = \sqrt{\frac{N_{++}^{\rm same} 
\cdot N_{--}^{\rm same}}{N_{++}^{\rm mix} \cdot N_{--}^{\rm mix}}}
\end{equation}
where $N_{++}^{\rm same}$, $N_{--}^{\rm same}$ are the like-sign yields in 
same events and $N_{++}^{\rm mix}$, $N_{--}^{\rm mix}$ are the 
like-sign yields in mixed events, for dimuon mass $\ge 2$ GeV$/c^{2}$. 
As shown in Fig.~\ref{fig:jpsi_mass}, the invariant mass 
distributions determined by these methods are consistent with each 
other within statistical uncertainties. The mixed event method is then 
used to determine the combinatorial background for the final analysis 
in order to reduce statistical fluctuations. After the combinatorial 
background subtraction, clear \jpsi peaks are found in both muon arms,
as shown in Figs.~\ref{fig:jpsi_mass}(c) and (d). A mass 
window cut ($2.7<\rm{mass}<3.5~\rm{GeV}/c^{2}$) is applied to the 
dimuon pair invariant mass distribution to select \jpsi candidates. The 
signal (combinatorial background subtracted yields) to the 
combinatorial background ratio in the \jpsi mass window is 18.6 in the 
$1.2<y<2.2$ region and 19.9 in the $-2.2<y<-1.2$ region.

\subsection{Simulation Setup}
\label{sec: Simulation Setup}

\begin{figure}[!htb]
	\includegraphics[width=1.0\linewidth]{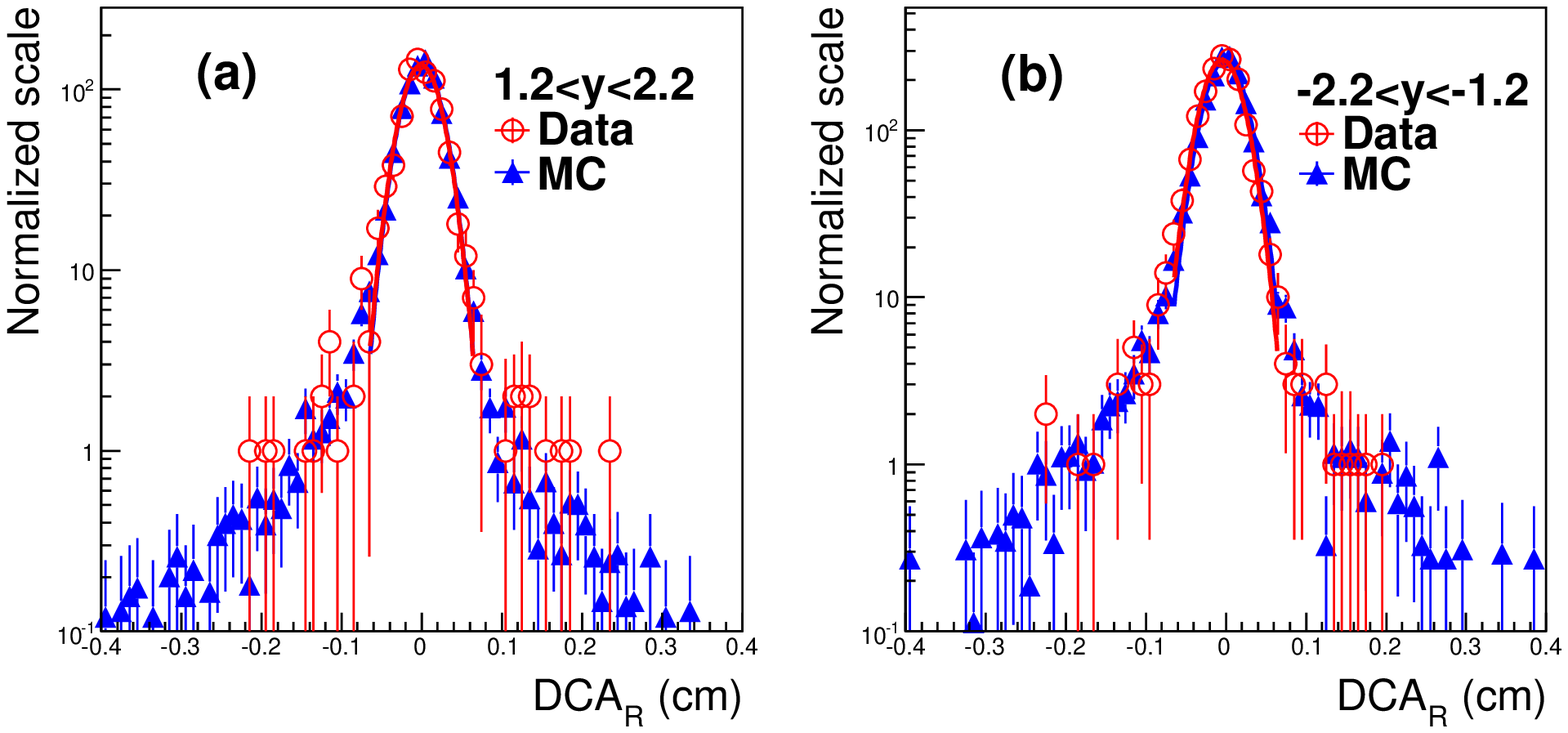}
	\includegraphics[width=1.0\linewidth]{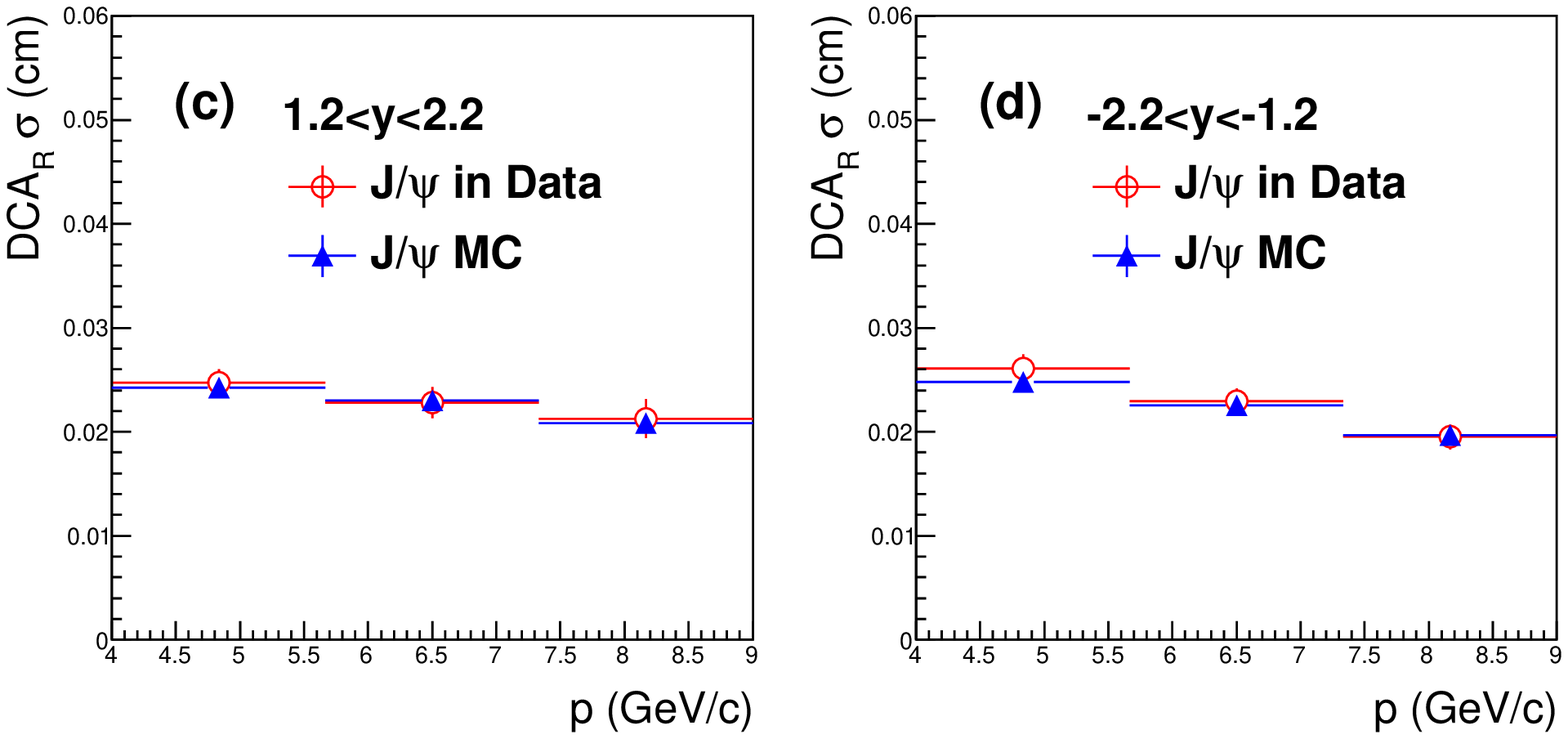}
\caption{\label{fig:dcar_dat_mc_comparison} Comparison of the normalized 
\dcar distributions of single muons from inclusive \jpsi events in data 
(red open circle) and simulation (blue solid triangle).  Panels (a) 
and (b) show the comparison for integrated momenta and 
panels (c) and (d) show the comparison for the momentum-dependent 
\dcar resolution. There is good agreement between data and simulation.}
\end{figure}

The full simulation framework, which comprises 
{\sc pythia8}\cite{Sjostrand:2007gs}+{\sc geant}4\cite{Agostinelli2003250}
+reconstruction, is set up to characterize the \dcar distributions of 
muons from prompt \jpsi and \jpsi from $B$-meson decay. Dead areas in 
the detector are determined from data on a run-by-run basis and the same 
vertex and tracking reconstruction algorithms as in data analysis are 
used. The width of the simulated primary vertex distributions along the 
$x$ and $y$ axes is 80 $\mu$m as determined from Vernier Scan 
measurements~\cite{Vernier}. The vertex distribution along the $z$ axis 
used in the simulation has been determined from the real data. To get 
an accurately reproduced $z$ vertex resolution in simulation, which is 
dependent on the multiplicity in the event, additional simulated MB 
events (with $z$ vertex matched to the hard QCD events) are embedded 
into the prompt \jpsi events, or events with a \jpsi from $B$-meson 
decay. To ensure that the accessed kinematic region of the probed 
parton distribution function (PDF) in the MB events is the same in 
prompt \jpsi events or in $B$-meson \rarr \jpsi events, the 
renormalization scale $Q^{2}_{\rm renorm}$ defined in {\sc pythia}, 
which determines the PDF shape, is kept at the same value between the 
MB event and the triggered event.

\begin{figure}[!htb]
	\includegraphics[width=0.95\linewidth]{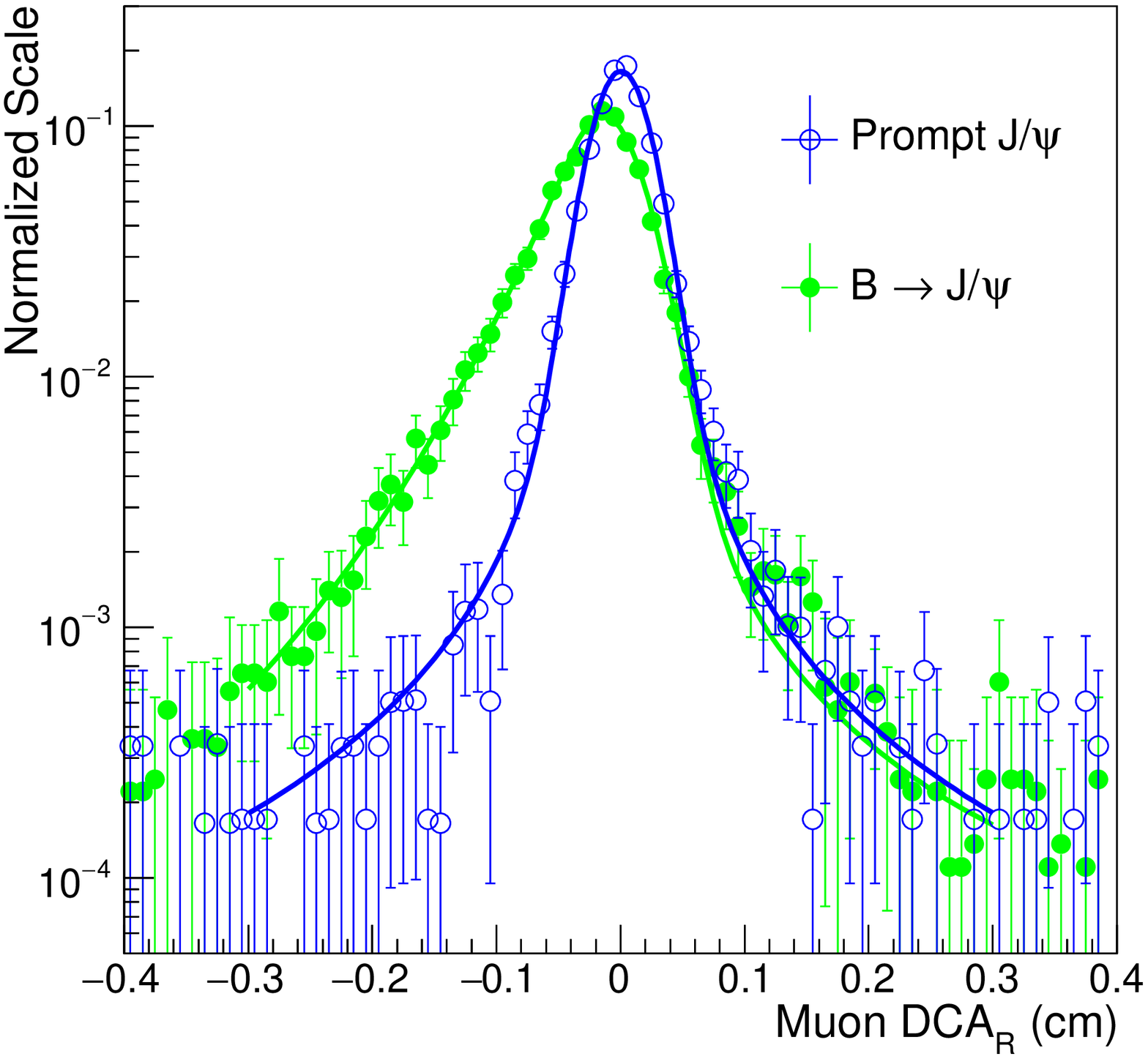}
\caption{\label{fig:mc_fit} Normalized \dcar distributions of simulated 
prompt \jpsi (blue open circle) and $B$-meson \rarr \jpsi events (green 
circle). The muon \dcar distributions are normalized by the total number 
of entries in the \dcar range of (-0.4cm, 0.4cm). Solid lines stand for 
the fits defined in Eq.~\ref{eq:jpsi_fun} and Eq. ~\ref{eq:bjpsi_fun}.}
\end{figure}

To verify that the simulations accurately represent the real data, 
we have compared the simulated and measured muon \dcar distributions from 
inclusive \jpsi events. The inclusive \jpsi events in 
simulation are obtained by combining 90\% prompt \jpsi events and 
10\% \jpsi from $B$-meson decay. This fraction of $B$-meson decays to 
\jpsi is selected based on the average result from global data measured 
in the same inclusive \jpsi $p_{T}$ region 
\cite{Acosta:2004yw,Abelev:2012gx,Khachatryan:2010yr,lhcb8TeV, 
Aaij:2015rla}. A single Gaussian function is fit to the centroid of the 
\dcar distributions in data and simulation to derive the resolutions of 
the prompt component of the $\rm{DCA}_{\rm R}$. The momentum dependence 
of this \dcar resolution extracted from the core region 
($|\rm{DCA}_{\rm R}|<500 \mu$m) is compared between data and simulation. 
As shown in Fig.~\ref{fig:dcar_dat_mc_comparison}, good agreement 
between data and simulation is achieved in both of the measured rapidity 
regions.

\subsection{Signal Determination}
\label{sec:Signal Determination}

The shapes of the \dcar distributions of muons from prompt \jpsi and 
those from $B$-meson \rarr \jpsi are characterized using the full 
simulation. Figure~\ref{fig:mc_fit} shows the resulting normalized 
distribution of \dcar for muons from prompt \jpsi events (blue open 
circle) and from $B$-meson \rarr \jpsi events (green circle). As 
explained at the end of Section~\ref{sec: Experiment setup}, the shape 
of the muon \dcar distribution in prompt \jpsi events is symmetric, 
which is consistent with expectations for prompt particle decays. The 
$\Lambda_{b}$, $B^\pm$, $B^{0}$, $B_{s}^{0}$ hadrons have a finite 
lifetime of 1.4--1.6 ps on average, resulting in a displaced vertex at 
forward rapidity of approximately 0.8 mm from the primary collision 
vertex for the \jpsi from $B$-meson decay. Due to the displacement 
between the decay vertex and the primary collision vertex, the negative 
side of the muon \dcar distribution shows a clear deviation from 
symmetry for $B$-meson \rarr \jpsi events. The respectively symmetric 
and asymmetric \dcar distributions allow the separation of prompt \jpsi 
from $B$-meson \rarr \jpsi .

Several functions were tested to describe the line shapes of the muon 
\dcar in both prompt \jpsi and \jpsi from $B$-meson decay in 
simulations. The final fit functions which will be described below are 
selected based on the best fits to the simulation spectra with the 
maximum log-likelihood method and the convolution of the intrinsic 
\dcar resolution with a function which represents $B$ meson decay 
kinematics is used. Variations of the fit functions and the simulation 
setup were then used to account for systematic uncertainties in the fit 
function. A convolution fit is used to describe the shape of the muon 
\dcar from prompt \jpsi decay, with the definition shown in Eq. 
(\ref{eq:jpsi_fun}).

\begin{eqnarray}
\label{eq:jpsi_fun}
f_{{\rm {prompt}} \ J/\psi}(\textrm{DCA}_{\rm R}) 
&=& \frac{1}{\sqrt{2\pi}\sigma}\textrm{exp}[-\frac{(\textrm{DCA}_{\rm R}-\mu)^{2}}{2\sigma^{2}}] \\\nonumber 
& \otimes & \frac{\sigma_{1}^{2}\textrm{DCA}_{\rm R}^{2}}{(\textrm{DCA}_{\rm R}^{2}-\mu_{1}^{2})^{2}+\textrm{DCA}_{\rm R}^{4}(\sigma_{1}^{2}/\mu_{1}^{2})} , 
\end{eqnarray}

\noindent where $\mu$, $\sigma$, $\mu_{1}$ and $\sigma_{1}$ are 
determined from the fit to the prompt \jpsi simulation spectra. 
Parameter $\sigma$ and $\sigma_{1}$ determine the width of the muon 
\dcar shape in prompt \jpsi events, which comes from the detector and 
vertex resolutions. Values of these parameters defined in Eq. 
(\ref{eq:jpsi_fun}) are fixed in the next step: the fit to the measured 
\dcar distributions. For $B$-meson decay to $J/\psi$ events, the 
convolution fit function defined in Eq. (\ref{eq:bjpsi_fun}) is used:

\begin{eqnarray}
\label{eq:bjpsi_fun}
f_{B \rightarrow J/\psi}(\textrm{DCA}_{\rm R}) 
&=& f_{{\rm {prompt}} \ J/\psi}(\textrm{DCA}_{\rm R}) \\\nonumber 
& \otimes &  f_{B}(\textrm{DCA}_{\rm R}) ,  
\end{eqnarray}

\noindent where the function $f_{{\rm {prompt}} \ 
J/\psi}(\textrm{DCA}_{\rm R})$ is defined in Eq. (\ref{eq:jpsi_fun}). 
The parameters of $f_{{\rm {prompt}} \ J/\psi}(\textrm{DCA}_{\rm R})$ 
are already determined, as explained above, in the fit of muon \dcar in 
the prompt \jpsi simulation. Function $f_{B}(\textrm{DCA}_{\rm R})$, 
which stands for the decay kinematics of $B$-meson, is defined as:

	\begin{equation}
	\label{eq:decay_B}
	f_{B}(\textrm{DCA}_{\rm R}) = 
	\left\{\begin{array}{ll} 
\textrm{exp}[-\frac{(\textrm{DCA}_{\rm R}-\mu_{2})^{2}}{2\sigma_{2}^{2}}],  
\frac{\textrm{DCA}_{\rm R} - \mu_{2}}{\sigma_{2}}>-\alpha \\
(\frac{n}{|\alpha|})^{n}\textrm{exp}(-\frac{|\alpha|^{2}}{2})(\frac{n}{|\alpha|}-|\alpha|-\frac{\textrm{DCA}_{\rm R}-\mu_{2}}{\sigma_{2}})^{-n},  
\frac{\textrm{DCA}_{\rm R} - \mu_{2}}{\sigma_{2}} \le -\alpha \\
	\end{array}\right.
	\end{equation}

\noindent where $\mu_2$, $\sigma_2$, $n$ and $\alpha$ are parameters 
determined from the fit to the 
$B\rightarrow J/\psi \rightarrow \mu^+\mu^-$ simulation. The average 
value of the muon \dcar from $B\rightarrow J/\psi$ decay is determined 
by $\mu_2$. Parameters $\sigma_2$, $n$ and $\alpha$ determine the 
asymmetric shape of this \dcar distribution. The determined values of 
these parameters defined in this section and used in 
Eq.~(\ref{eq:bjpsi_fun}) and Eq.~(\ref{eq:decay_B}) are then fixed in the 
fit to the measured \dcar distributions.

Fits of the simulated muon \dcar distributions for prompt $J/\psi$ (blue 
open circle) and $B$ to $J/\psi$ (green circle) are shown in 
Fig.~\ref{fig:mc_fit}. The \dcar spectra can be modeled by the two functions 
defined in Eq. (\ref{eq:jpsi_fun}) and Eq. (\ref{eq:bjpsi_fun}). 

\begin{figure}[!htb]
	\includegraphics[width=0.96\linewidth]{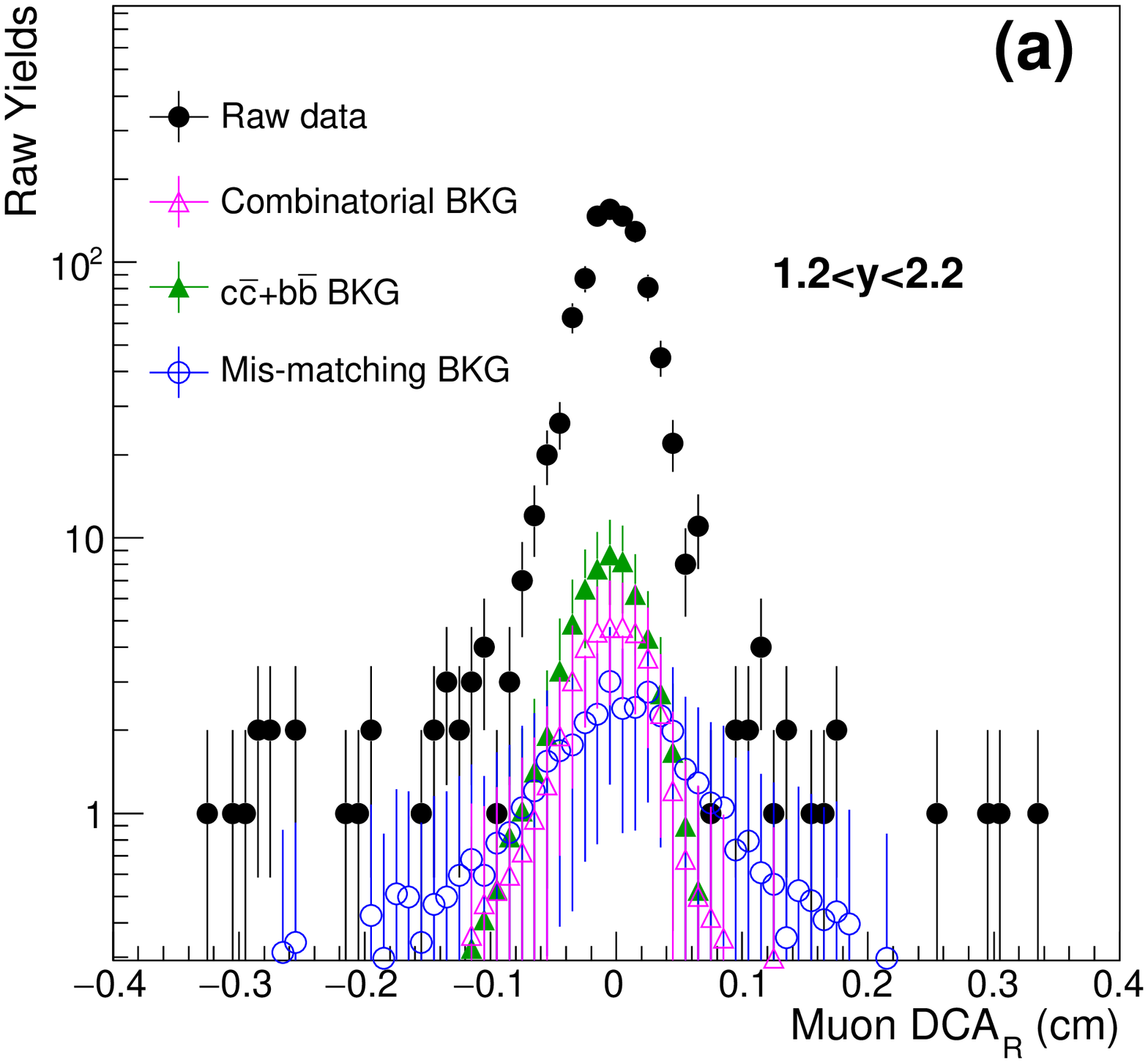}
	\includegraphics[width=0.96\linewidth]{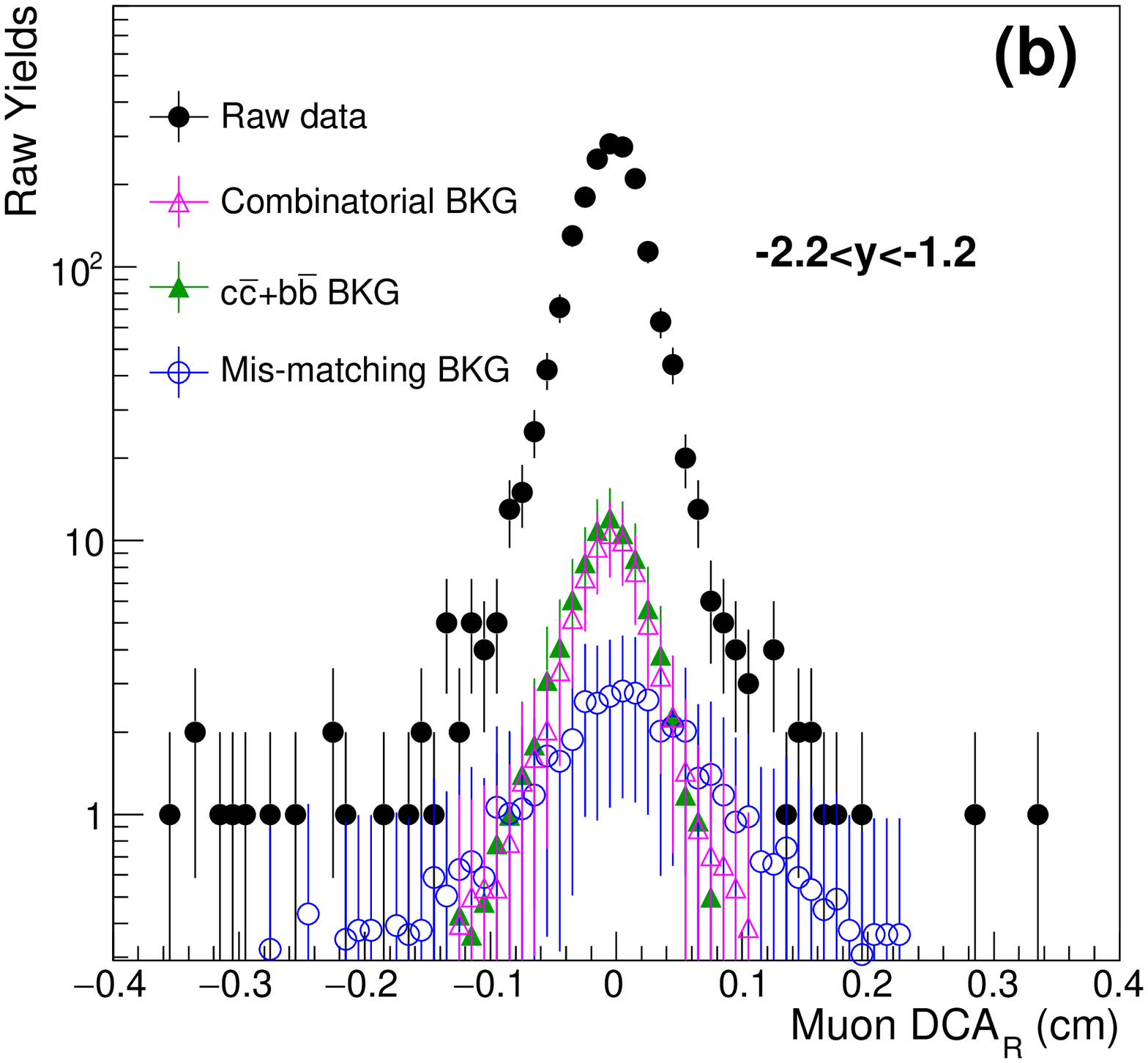}
\caption{\label{fig:bkg_dis} The raw yields of data ([black] closed 
circles) and estimated background \dcar distributions within the \jpsi 
mass window (2.7--3.5 GeV/$c^{2}$ are shown for (a) rapidity $1.2<y<2.2$ 
and (b) $-2.2<y<-1.2$ The combinatorial background defined as $f_{\rm 
combinatorial}$ in Eq. (\ref{eq:bkgdis}) ([magenta] open triangle), the 
heavy flavor continuum ($c\bar{c}+b\bar{b}$) background defined as 
$f_{c\bar{c}+b\bar{b}}$ in Eq. (\ref{eq:bkgdis}) ([green] solid 
triangle) and the detector mis-matching background defined as $f_{\rm 
mismatch}$ in Eq. (\ref{eq:bkgdis}) ([blue] open circle) are determined 
using techniques described in the text.}
\end{figure}

\subsection{Background Determination}
\label{sec:Background Determination}

For this analysis, backgrounds come from three different sources: 
combinatorial, MuTr-FVTX track mis-matching and heavy flavor decay 
continuum which represents unlike-sign dimuon pairs from 
$b\bar{b} \rightarrow B\bar{B} \rightarrow \mu^{+}\mu^{-} + X$ and 
$c\bar{c} \rightarrow D\bar{D} \rightarrow \mu^{+}\mu^{-} + X$ events.
The combinatorial background and the background from 
mis-matching between FVTX and MuTr tracks are determined by data-driven 
methods. The fraction of the contribution from the heavy flavor 
continuum background is determined by fitting the dimuon pair invariant 
mass spectra in real data, and the \dcar shape is determined from 
simulation. Details of the background determinations will be discussed 
in sections~\ref{sec:comb_background} through~\ref{sec:hf_background}.

\subsubsection{Combinatorial Background Determination}
\label{sec:comb_background}

The combinatorial background, which comes from combining randomly 
associated tracks in an event, is evaluated using unlike-sign dimuons 
formed by muon tracks from two different events (referred to as the 
mixed event procedure). The events to be mixed are required to have $z$ 
vertices with no more than 1.5 cm difference from each other. The muon 
\dcar distribution of the combinatorial background from normalized mixed 
events (the normalization factor is defined in Eq. (\ref{eq:norm_mix})) 
is shown as magenta open triangles in Fig.~\ref{fig:bkg_dis}.

\begin{figure}[!htb]
	\includegraphics[width=0.96\linewidth]{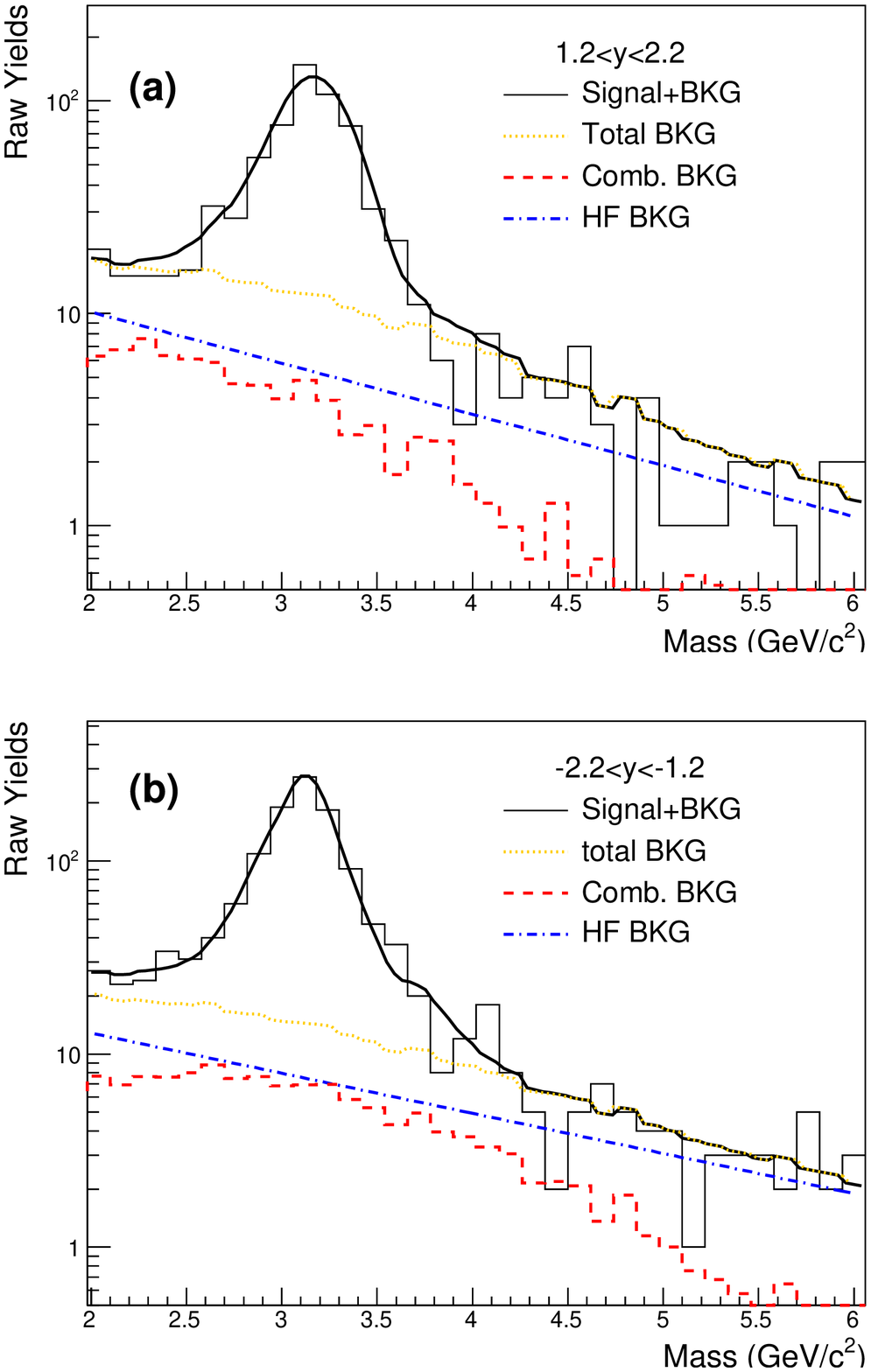}
\caption{\label{fig:dimu_mass_fit} Fit of dimuon mass spectra to 
determine the heavy flavor continuum background for (a) rapidity 
$1.2<y<2.2$ and (b) $-2.2<y<-1.2$.  The fit function ([black] solid 
curve) includes the \jpsi and $\psi \prime$ yields which already 
include the FVTX-MuTr mismatching background, the combinatorial 
background ([red] dashed curve) and the heavy-flavor-continuum 
background ([blue] dash-dotted curve). The total background 
([yellow] solid-dotted curve) shows the combinatorial and the 
heavy-flavor-continuum background.}
\end{figure}

\subsubsection{FVTX-MuTr mis-matching determination}
\label{sec:mismatch}

The last FVTX plane and the first MuTr station are 150 cm apart and have 
approximately 1 m of absorber material in between. MuTr tracks with 
momentum above 3 GeV/$c$ projected to the fourth station of the FVTX 
therefore cover a circle with a radius of up to 2 cm for muons, due to 
the multiple scattering in the absorber. As a result, some fraction of 
the MuTr projections will find more than one FVTX track or a single but 
incorrect FVTX track inside its projected circle, and have a certain 
probability of selecting an incorrect FVTX match. We refer to these 
incorrect matches as ``mis-matching background''.

To estimate the amount of mis-matching, we attempt to match MuTr tracks 
from one event to FVTX tracks from a separate event (referred to as 
swapped events). To be as realistic as possible, the swapped events need 
to belong to the same $z$-vertex category, meaning the difference of the 
$z$ vertex between the swapped event and the true event should be less 
than 1mm. The selection of 1mm $z$-vertex difference does not introduce 
any bias to the \dcar distribution. In addition to this, we also count 
the mis-matching tracks from swapped events only when the matching track 
in the swapped event has a better $\chi^{2}$ than the matching track in 
the real event, so that we do not overestimate the mismatches in real 
events. The mis-matching background in the analyzed events is dominated 
by \jpsi MuTr tracks which do not have a corresponding FVTX track in the 
real event and accidentally match to a random background track. The 
fraction of candidate FVTX tracks in swapped events which are found to 
be wrongly associated with a MuTr track from a good \jpsi dimuon pair, 
and that pass the quality cuts shown in Table~\ref{tab:quality_cut}, is 
3\% (2\%) in the $1.2<y<2.2$ ($-2.2<y<-1.2$) rapidity region.

\subsubsection{Heavy Flavor Background Determination}
\label{sec:hf_background}

After subtracting the combinatorial background from the dimuon invariant 
mass distribution within the 2--6 GeV$/c^{2}$ region (shown in 
Fig.~\ref{fig:jpsi_mass}), there are remaining backgrounds in the sideband 
regions outside the $J/\psi$ mass window. This remaining background is 
dominated by the heavy flavor continuum and indicates that this 
continuum is not negligible in the $J/\psi$ mass region. To determine 
the fraction of the heavy flavor background, a fit function which 
includes yields from $J/\psi$, $\psi'$, the combinatorial background and 
heavy flavor continuum background is applied to the invariant mass 
distribution of dimuon pairs. In the dimuon pair mass region $>4$ 
GeV$/c^{2}$, the heavy flavor continuum background also contains 
Drell-Yan. Because the fraction of Drell-Yan events within the \jpsi mass 
region (2.7--3.5 GeV$/c^{2}$) is negligible, the fit in this mass 
region does not include a Drell-Yan component.

Figure~\ref{fig:dimu_mass_fit} shows the fit of the dimuon mass 
distribution to determine the heavy flavor continuum background. The 
total background (yellow) determined by the fit to the invariant mass 
spectrum, which comprises the combinatorial (red) and the heavy flavor 
background (blue), follows the mass distribution outside the \jpsi mass 
window well. The fraction of the heavy flavor background within the 
$J/\psi$ mass window is found to be 7.1\% $\pm$ 1.1 \% (5.5\% $\pm$ 0.8\%) in the $1.2<y<2.2$ 
($-2.2<y<-1.2$) regions.

The relative $b\bar{b}$ and $c\bar{c}$ dimuon contributions within the 
\jpsi mass window are not well known, and extrapolation from previous 
midrapidity dimuon invariant mass yields in 200 GeV $p$+$p$ collisions 
would introduce a large systematic uncertainty. We therefore first fit 
the unlike-sign dimuon invariant mass spectrum near the \jpsi region 
including the {\sc pythia}8-simulated shape of $b\bar{b}$ and $c\bar{c}$ 
components and an unconstrained normalization scale to estimate the 
contribution. The fit suggests there is a 33\% $b\bar{b}$ fraction in 
the heavy flavor continuum within the \jpsi mass region. However, we do 
note there is systematic uncertainty in the {\sc pythia}8 shape. Because of 
this uncertainty, for this analysis the fraction of the $b\bar{b}$ 
contribution to the heavy flavor yields within the \jpsi mass window is 
set to be 50\%, and varied from 0 to 100\% to take into account all 
possibilities in the systematic uncertainty.

\subsection{Fitting Procedure}
\label{sec: Fitting Procedure}

\begin{figure*}[]
	\includegraphics[width=0.48\linewidth]{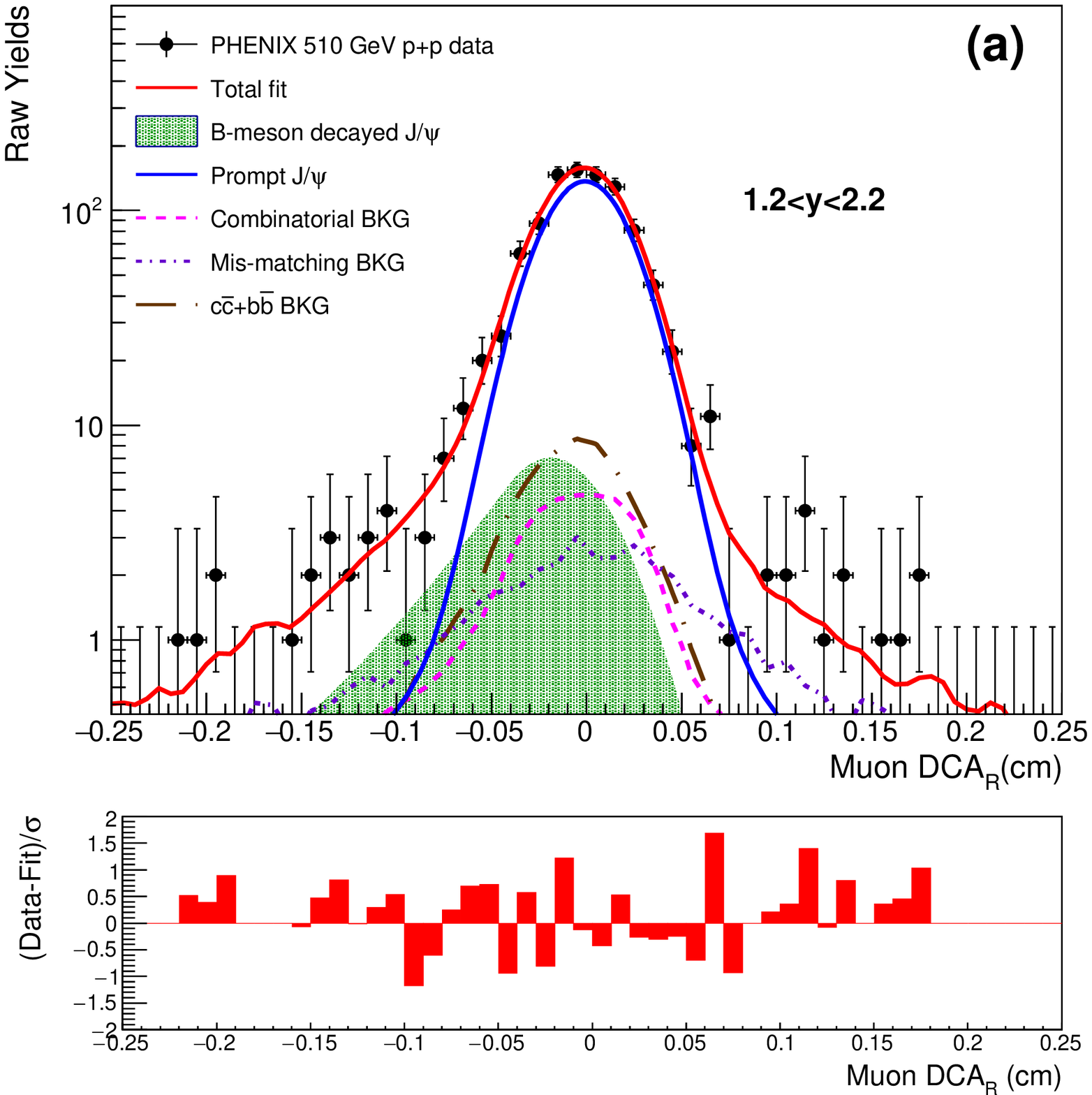}
	\includegraphics[width=0.48\linewidth]{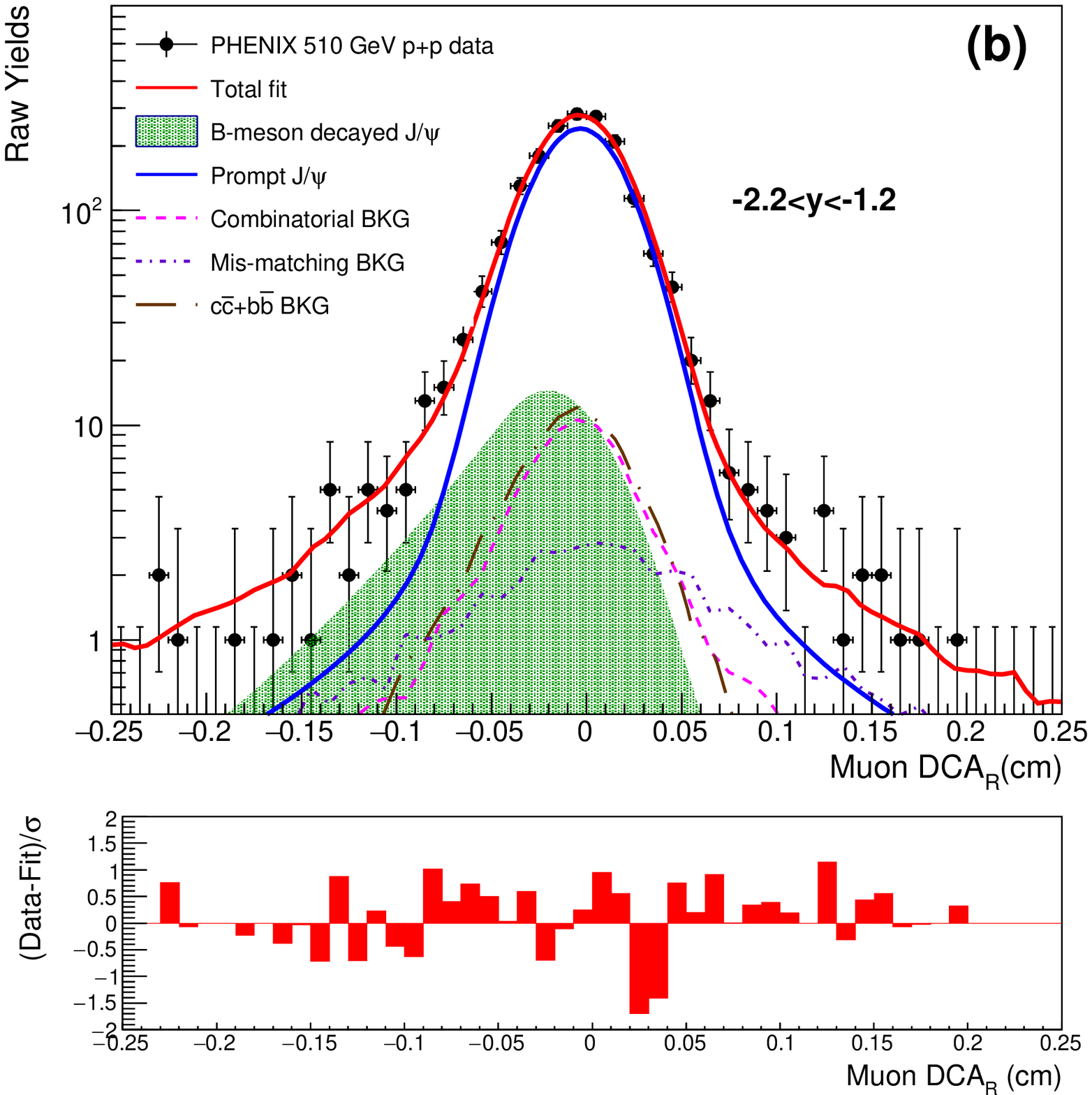}
\caption{\label{fig:dcar_fit} $B{\rightarrow}$\jpsi fraction fit to muon 
\dcar in the (a) $1.2<y<2.2$ and (b) $-2.2<y<-1.2$ regions. The 
([red] solid curve) stands for the total fit, which includes the 
prompt \jpsi (solid blue), the $B$-meson \rarr \jpsi ([green] filled 
region), the combinatorial background ([magenta] dashed curve), 
the $c\bar{c}+b\bar{b}$ background ([brown] long-dashed curve) 
and the detector mismatching background ([purple] short-dashed curve).}
\end{figure*}

The \dcar distributions are selected from dimuon pairs within the mass 
window 2.7--3.5 GeV/$c^{2}$. A fit function is developed to 
simultaneously extract the prompt \jpsi and $B$-meson \rarr \jpsi 
yields from the real data \dcar distributions with the maximum 
log-likelihood method.
This fit function comprises five components: 1) muons from prompt 
$J/\psi$, 2) muons from $B$-meson \rarr $J/\psi$, 3) combinatorial 
background determined by mixed events, 4) mismatching between FVTX and 
MuTr determined by swapped events, and 5) heavy flavor 
($c\bar{c}+b\bar{b}$) continuum background. 
The fit function which is used to determine the shape of muon \dcar 
distributions from prompt \jpsi ($B$-meson \rarr $J/\psi$) events is 
$f_{\textrm{prompt} \ J/\psi}(\textrm{DCA}_{\rm R})$ ($f_{B \rightarrow 
J/\psi}(\textrm{DCA}_{\rm R})$) as discussed in Section~\ref{sec:Signal 
Determination}. Parameters defined in both Eq. (\ref{eq:jpsi_fun}) and 
Eq. (\ref{eq:bjpsi_fun}) are fixed according to the fit to the simulated 
spectra and the detector resolution smearing is fine-tuned in the data 
fit. The functions which represent the three background contributions 
are $f_{\rm combinatorial}(\textrm{DCA}_{\rm R})$, $f_{\rm 
mismatch}(\textrm{DCA}_{\rm R})$ and 
$f_{c\bar{c}+b\bar{b}}(\textrm{DCA}_{\rm R})$ as discussed in Section 
\ref{sec:Background Determination}. Histograms of muon \dcar from 
different background contributions after normalization are used to 
represent each component in Eq. (\ref{eq:bkgdis}). Fluctuations of the 
fit methods, signal and background determinations are studied in the 
systematic uncertainty evaluations. These functions used to describe 
the data spectrum, are summarized in Eq. (\ref{eq:total_fun}),

\begin{align}
\label{eq:total_fun}
   f_{\rm total}(\textrm{DCA}_{\rm R}) 
& =  f_{\rm sig}(\textrm{DCA}_{\rm R}) 
+ f_{\rm bkg}(\textrm{DCA}_{\rm R}) , \\
\label{eq:sig_fun}
   f_{\rm sig}(\textrm{DCA}_{\rm R}) 
& =  \rm{Yield}_{\rm incl. \jpsi} \times  \\\nonumber                                                   
&  [ \textrm{F}_{B{\rightarrow}J/\psi} 
\times f_{B \rightarrow J/\psi}(\textrm{DCA}_{\rm R}) \\\nonumber                                                   
& +  (1-\textrm{F}_{B{\rightarrow}J/\psi}) 
\times f_{\textrm{prompt} \ J/\psi}(\textrm{DCA}_{\rm R})] , \\
 \label{eq:bkgdis} 
   f_{\rm bkg}(\textrm{DCA}_{\rm R}) 
& =  f_{\rm combinatorial}(\textrm{DCA}_{\rm R}) \\\nonumber                                                 
& +  f_{\rm mismatch}(\textrm{DCA}_{\rm R}) \\\nonumber                                                 
& +  f_{c\bar{c}+b\bar{b}}(\textrm{DCA}_{\rm R}) ,  \\\nonumber
\end{align}

\noindent where $\rm{Yield}_{\rm incl\ \jpsi}$ is the total yield of 
inclusive \jpsi which comprises both prompt \jpsi and $B$-meson decayed 
\jpsi. Normalization and shapes of most of the components are fixed in 
previous steps. In the final stage of the fit, the fraction of muons 
from $B$-meson \rarr \jpsi (i.e. \bfrac), is the main free parameter in 
the total fit function (defined in Eq. (\ref{eq:total_fun})), together 
with the \jpsi yield and a last tuning of the resolution that is 
described below. As the \dcar resolution in data can be affected by 
additional factors which may not be well captured by the simulation 
(such as event-by-event variations in the vertex resolution, additional 
smearing from multiple scattering in the nonuniform detector materials, 
part of the detector randomly dropping out within a run and beam-beam 
collision geometry fluctuations), an additional free parameter, $\sigma 
\prime$, is introduced in the convolution fit functions for prompt 
\jpsi (defined in Eq.  (\ref{eq:jpsi_fun})) and $B$-meson \rarr 
$J/\psi$ (defined in Eq.  (\ref{eq:bjpsi_fun})). It accounts for 
detector resolution smearing and also captures any uncertainty of the 
beam spot size. The fit is then performed with the parameter 
$\sigma_{1}^{*}$ instead of $\sigma_{1}$, where $\sigma_{1}^{*} = 
\sigma_{1} + \sigma \prime$. The resolution smearing parameter $\sigma 
\prime$ determined from the fit to the data is within 20 $\mu$m with 
approximately 20 $\mu$m statistical uncertainty for the 1.2 $<|y|<$ 2.2 
region. The size of the smearing is much smaller than the the average 
$x$-$y$ beam profile value (around 80 $\mu$m) and the \dcar resolution 
(around 230 $\mu$m). The value of the resolution smearing $\sigma 
\prime$ varies from 5 to 70 $\mu$m when different beam profile values 
in the $x$-$y$ plane are used in the simulation (from 80 to 180 
$\mu$m). Variation of the smearing parameter $\sigma \prime$ will be 
included in the systematic uncertainty evaluation. Applying the fit 
procedure to the \dcar distributions, assuming 50\% of the heavy flavor 
continuum contribution comes from $b\bar{b}$ (see discussions in 
\ref{sec:hf_background}), allows the raw fraction of \jpsi mesons from 
$B$ decays in inclusive \jpsi yields to be extracted. The corresponding 
raw ratios $B \rightarrow$ \jpsi are $7.3\% \pm 3.7\%$ (stat) for (1.2 
$<y<$ 2.2) and $8.1\% \pm 2.8\%$ (stat) for (-2.2$<y<$-1.2). The 
spectra and fit results are shown in Fig.~\ref{fig:dcar_fit}. The fit 
parameter values are summarized in Table~\ref{tab:fit_par}.

\begin{table}[!htb]
\caption{\label{tab:fit_par} Parameters as defined in 
Eq.(\ref{eq:jpsi_fun}) and Eq. (\ref{eq:bjpsi_fun}). Most of these 
parameters are fixed in preliminary steps. At the final fit stage, free 
parameters are $B{\rightarrow}$\jpsi fraction (\bfrac) (see text), the 
total \jpsi yields and $\sigma \prime$. Uncertainties are not only from 
the statistical fluctuations but also related with the systematic 
uncertainty evaluations.}
\begin{ruledtabular} \begin{tabular}{lccc}  
    Fit parameter & \ -2.2 $<y<$ -1.2 \ & \ 1.2 $<y<$ 2.2 \\\hline
          $\mu$       &   -15 $\pm$ 5 $\mu$m     &    6 $\pm$ 5 $\mu$m \\
         $\sigma$    &    209 $\pm$ 8 $\mu$m     &   210 $\pm$ 6 $\mu$m  \\
         $\mu_{1}$   &     0 $\mu$m    &      0 $\mu$m   \\
         $\sigma_{1}$   &  60 $\pm$ 11 $\mu$m     &   50 $\pm$ 9$\mu$m      \\
         $\sigma \prime$ &    7 $\pm$ 14 $\mu$m   &   10 $\pm$ 18 $\mu$m   \\
         $\mu_{2}$   &    -135  $\pm$ 15 $\mu$m    &    -123 $\pm$ 18 $\mu$m   \\
         $\sigma_{2}$ &   169 $\pm$ 10 $\mu$m    &    150 $\pm$ 16 $\mu$m \\
         $\alpha$      &   0.74 $\pm$ 0.06     &   0.60 $\pm$ 0.08   \\
         $n$      &     3.50 $\pm$ 0.51  &   4.26 $\pm$ 0.75  \\
      \end{tabular} \end{ruledtabular} 
\end{table}

\subsection{Acceptance$\times$Efficiency Correction}
\label{sec: correction}

In \pp collisions, the \dcar resolution is dominated by the VTX/FVTX 
vertex resolution. Higher event multiplicity can lead to a better vertex 
resolution and a higher probability that a vertex can be reconstructed 
for a given event. The $B{\rightarrow}J/\psi$ events have higher average 
VTX/FVTX multiplicity in comparison with prompt \jpsi events. 
Conversely, due to their different $p_{T}$ distributions, $B \rightarrow 
J/\psi$ events have a somewhat lower probability of having both muons 
accepted into the muon arm than prompt \jpsi events. These differences 
in VTX/FVTX event multiplicities and kinematics result in somewhat 
different values of the acceptance$\times$efficiency for the two sets of 
events. The raw ratio $F_{B{\rightarrow}J/\psi}^{\rm raw}$ as discussed in 
section~\ref{sec: Fitting Procedure} must be corrected for the relative 
acceptance$\times$efficiency difference between prompt \jpsi and 
$B{\rightarrow}J/\psi$ events, using the 
{\sc pythia}8+{\sc geant}4+reconstruction simulation 
described previously in section~\ref{sec: Simulation Setup}, 
$\frac{A\varepsilon_{{\rm prompt}~J/\psi{\rightarrow}\mu\mu}}
{A\varepsilon_{B{\rightarrow}J/\psi{\rightarrow}\mu\mu}}$, where 
$A\varepsilon_{{\rm prompt}~J/\psi{\rightarrow}\mu\mu}$ 
($A\varepsilon_{B{\rightarrow}J/\psi{\rightarrow}\mu\mu}$) is the 
acceptance$\times$efficiency for prompt \jpsi ($B{\rightarrow}J/\psi$) events.
 
The acceptance$\times$efficiency for prompt \jpsi events is 0.455\% $\pm$ 0.007\% 
(0.506\% $\pm$ 0.008\%) and for $B{\rightarrow}$\jpsi events is 0.446\% $\pm$ 0.007\% (0.473\% $\pm$ 0.007\%) in the 
$-2.2<y<-1.2$ ($1.2<y<2.2$) rapidity region. The extracted relative 
ratio of $B{\rightarrow}$\jpsi acceptance$\times$efficiency to prompt \jpsi 
acceptance$\times$efficiency is 0.980 $\pm$ 0.022 (0.935 $\pm$ 0.020) in the $-2.2<y<-1.2$ 
($1.2<y<2.2$) rapidity region. The $B{\rightarrow}J/\psi$ fraction 
$F_{B{\rightarrow}J/\psi}$ which is defined as $\frac{N_{B{\rightarrow}J/\psi}}{N_{{\rm 
{prompt}}~J/\psi} + N_{B{\rightarrow}J/\psi}}$ ($N_{{\rm 
{prompt}}~J/\psi}$ is the yield for prompt $J/\psi$, 
$N_{B{\rightarrow}J/\psi}$ is the yield for $B{\rightarrow}J/\psi$) can be 
derived according to Eq. (\ref{eq:eff_corr3}).

\begin{equation}
\label{eq:eff_corr3}
F_{B{\rightarrow}J/\psi} 
= \frac{1}{1+(\frac{1}{F_{B{\rightarrow}J/\psi}^{\rm raw}}-1) 
\cdot \frac{\varepsilon_{B{\rightarrow}J/\psi{\rightarrow}\mu\mu}}
{\varepsilon_{{\rm {prompt}}~J/\psi{\rightarrow}\mu\mu}}}
\end{equation}

\subsection{Systematic Uncertainty}
\label{sec:Systematic Error}

The systematic uncertainty for $F_{B{\rightarrow}J/\psi}$ is evaluated 
by taking into account any factors which can affect the \dcar mean, the 
\dcar resolution, or the overall normalization of the signals. The 
following items are considered in the systematic uncertainty 
evaluation, along with a description of the methods performed to 
extract the uncertainties. For each item we compare the nominal $B\to$ 
\jpsi fraction extracted from our analysis to that obtained with 
alternate methods to extract the systematic uncertainty:

\begin{description}

\item [a] $p_{T}$ uncertainties: the $B$-meson \rarr \jpsi $p_{T}$ 
distributions were re-weighted in $B\to$ \jpsi simulations according to 
the prompt \jpsi $p_{T}$ distribution. The inclusive \jpsi $p_{T}$ 
spectrum was also varied with different fractions of prompt \jpsi and 
$B$-meson \rarr \jpsi.

\item [b] Background determination uncertainties: smooth fit functions 
were used to characterize the combinatorial, mismatching and heavy 
flavor backgrounds instead of histograms and their effect on the fit 
result was evaluated.

\item [c] Background determination uncertainties: deviation of fit 
results from the average value with different fractions of $b\bar{b}$ 
contribution in the heavy flavor background. The $b\bar{b}$ fraction of 
the heavy flavor background is varied from 0, 50\% to 100\%. Even 
though the assumption of 0 or 100\% $b\bar{b}$ heavy flavor continuum 
background is unrealistic, to be conservative, the maximum variation 
between the average value of the fitted $B$ to $J/\psi$ fraction and the 
fit result assuming 0 or 100\% $b\bar{b}$ fraction of heavy flavor 
background is quoted as the systematic uncertainty.

\item [d] Background determination uncertainties: the combinatorial 
background normalization Norm$_{\rm mix}$ defined in Eq.  
(\ref{eq:norm_mix}) was calculated within different dimuon mass ranges 
and compared to the nominal values.

\item [e] Fitting method uncertainties: multiple tests of the \dcar 
fit function with varied \dcar means and resolutions were applied to 
pseudo data, including different fractions of prompt \jpsi and \jpsi 
from $B$-meson decay with muon \dcar shape determined in simulation and 
realistic backgrounds. The stability of the extracted ratios was checked 
and deviation from the average value is accounted for in the systematic 
uncertainty.

\item [f] Signal determination uncertainties: different functions were 
used to represent the muon \dcar distributions in both prompt \jpsi and 
\jpsi from the $B$-meson decay events in simulation. A triple Gaussian 
function was used for prompt \jpsi events and a Crystal-Ball plus single 
Gaussian function was used for \jpsi from the $B$-meson decay events. 
The stability of the extracted ratios was checked.

\item [g] \jpsi selection uncertainties: good \jpsi candidates were 
selected in different dimuon pair mass windows (shifted by 0.15 
GeV/$c^{2}$) and the extracted ratio results were compared to the 
nominal ratios.

\item [h] Alignment determination uncertainties: different 
misalignment residuals were applied to the \dcar mean to determine their 
effect on the fit.

\item [i] Event quality cut uncertainties: different vertex resolution 
cuts were used and their effect on the fit evaluated.

\item [j] Dependence of simulation on different $x$-$y$ vertex 
smearing: the vertex smearing was varied from the reconstructed value in 
real data (around 200 $\mu$m) to the average beam profile value (around 
80 $\mu$m) and the effect on the fit evaluated.

\item [k] Variation of the acceptance$\times$efficiency: the 
renormalization scale factors were varied in simulation to get different 
$p_{T}$ distributions for prompt \jpsi and $B$ meson decays, then the 
acceptance$\times$efficiency correction factors were re-calculated and 
their effect on the fit was evaluated.

\end{description}

\begin{table*}
\caption{\label{tab:sys summary} Systematic uncertainty summary for the 
fraction of \jpsi from $B$-meson decay in the $1.2<y<2.2$ and 
$-2.2<y<-1.2$ rapidity regions. Values are in absolute scale. See the 
specific meaning of each item in Section~\ref{sec:Systematic Error}.}
\begin{ruledtabular}  \begin{tabular}{cccp{0.9\linewidth}}
Source   &    $1.2<y<2.2$ \  \  &  \  $-2.2<y<-1.2$ \ \  &  \  \ Specific meaning \\ \hline 
\\[-1em]
a     &   $<0.1\%$      &    $<0.1\%$   &   $p_{T}$ uncertainties.  \\  
b     &      0.1\%     &     0.2\%   &   Backgrounds shape variations with fit functions.  \\  
c     &    1.4\%       &    1.1\%   &   $b\bar{b}$ fraction variations in the heavy flavor background.  \\   
d     &     $<0.1\%$    &     $<0.1\%$   &   Combinatorial background normalization variation.  \\  
e      &     0.5\%      &    0.5\%   &   Fit method variations.  \\ 
f     &     $0.3\%$      &    0.3\%   &   Signal determination variations.   \\ 
g     &     0.4\%      &       0.5\%   &   $J/\psi$ selection variation.  \\
h      &      0.3\%     &        0.5\%   &   Alignment correction variations.   \\
i     &     0.4\%       &       0.6\%   &   Event quality cut variations.  \\
j     &     1.0\%       &       1.0\%   &   Vertex smearing in the $x-y$ plane.  \\
k    &     0.1\%       &        0.2\%   &   Variations of the acceptance$\times$efficiency.  \\
\\[-1em]
\textbf{Total syst uncertainty}   &     \textbf{1.9\%}     &        \textbf{1.9\%}         \\
\end{tabular} \end{ruledtabular}
\end{table*}

Table~\ref{tab:sys summary} gives the values and specific meanings for
each evaluated contribution to the systematic 
uncertainty on the extracted fraction for \jpsi from $B$-meson decay.  
As indicated, the total systematic 
uncertainty is 1.9\% in absolute scale for each muon arm in the 
$1.2<|y|<2.2$ rapidity coverage.

\begin{figure*}[!htb]
    \includegraphics[width=0.48\linewidth]{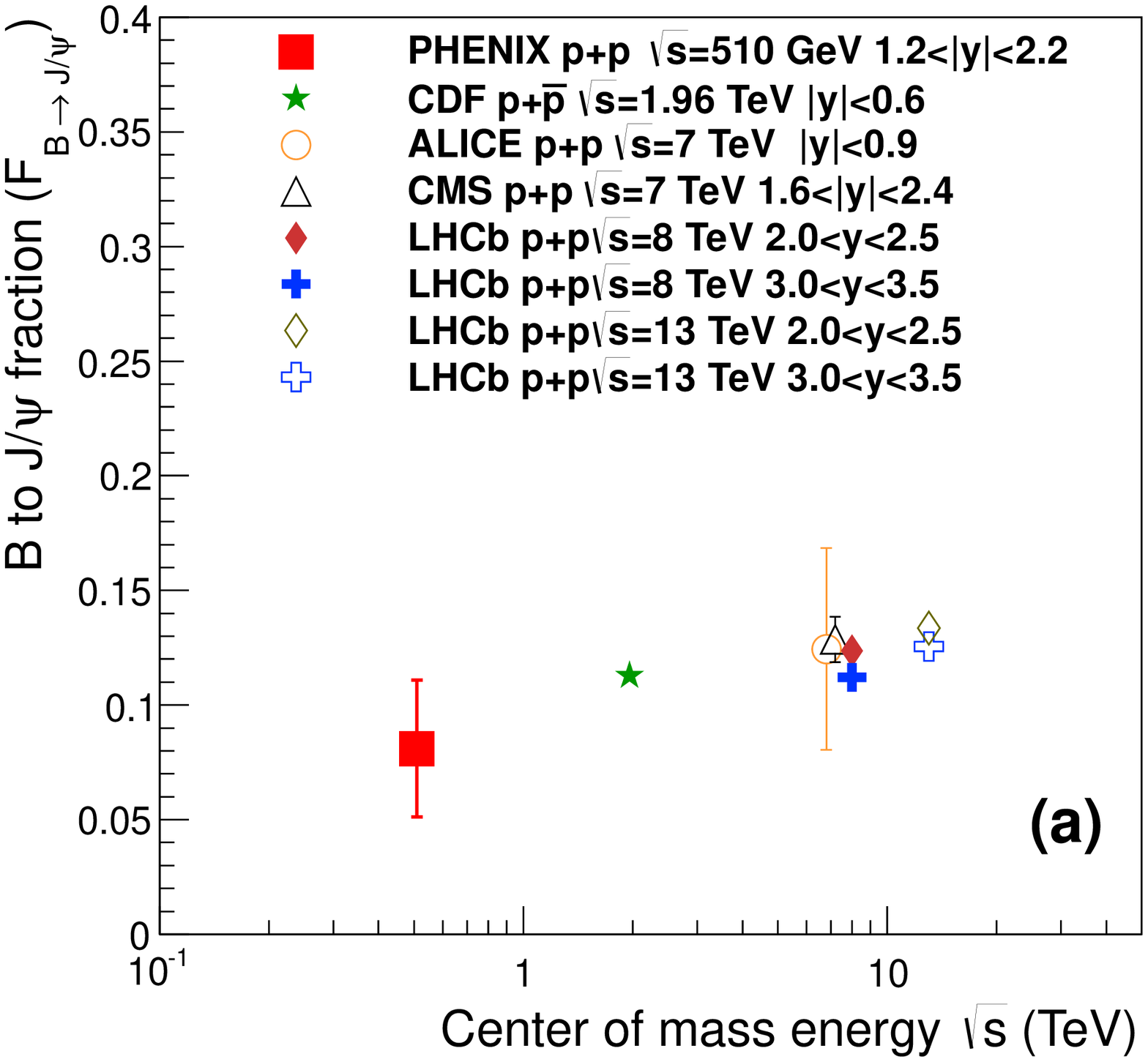}
    \includegraphics[width=0.48\linewidth]{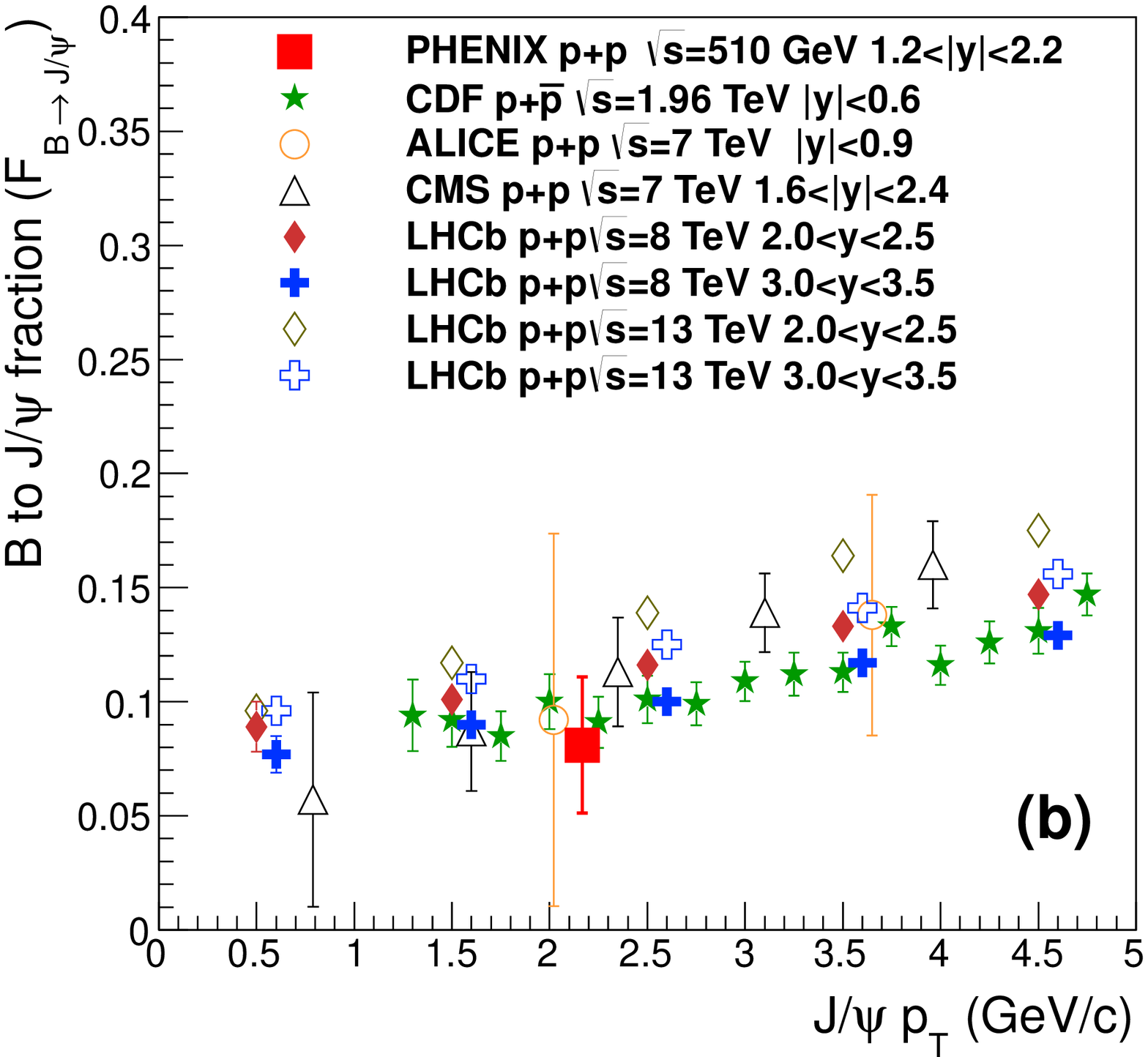}
\caption{\label{fig:global_data} Comparison of PHENIX $B$ $\rightarrow$ 
$J/\psi$ fraction with the global data from CDF~\cite{Acosta:2004yw}, 
ALICE~\cite{Abelev:2012gx}, CMS~\cite{Khachatryan:2010yr} and LHCb 
\cite{lhcb8TeV, Aaij:2015rla} experiments for \jpsi $p_{T}$ range of 
$0<p_{T}<5$ \gevc, (a) as a function of center of mass energy integrated 
in the \jpsi $0<p_{T}<5$ GeV/$c$ interval, and (b) as a function of 
inclusive \jpsi $p_{T}$.  The uncertainty of the PHENIX measurement 
is statistical and systematic combined.}
\end{figure*}

\section{Results and Discussions}
\label{sec: results_and_discussions}

After applying the acceptance$\times$efficiency factors shown in Table 
\ref{tab:acceptance}, the corrected $B{\rightarrow}$\jpsi fraction in the 
rapidity interval ($1.2 <y< 2.2$) is $7.8\% \pm 3.9\%$ (stat) and the 
fraction in the rapidity interval ($-2.2 <y< -1.2$) is $8.3\% \pm 2.9\%$ 
(stat).

\label{sec:rel_acceptance}

\begin{table} [!htb]
\caption{\label{tab:acceptance} Relative ratio of 
acceptance$\times$efficiency between prompt \jpsi and $B{\rightarrow}$\jpsi 
events, uncorrected $B{\rightarrow}$\jpsi fraction 
($F_{B{\rightarrow}J/\psi}^{\rm raw}$) and corrected $B{\rightarrow}$\jpsi 
fraction ($F_{B{\rightarrow}J/\psi}$). Uncertainties are statistical 
only.}
\begin{ruledtabular} \begin{tabular}{lccc}
& $\frac{A\varepsilon_{B{\rightarrow}J/\psi{\rightarrow}\mu\mu}}
{A\varepsilon_{{\rm {prompt}}~J/\psi{\rightarrow}\mu\mu}}$ 
& $F_{B{\rightarrow}J/\psi}^{\rm raw}$ & $F_{B{\rightarrow}J/\psi}$ \\ \hline 
-2.2 $<y<$ -1.2  & 0.980 $\pm$ 0.022 & $8.1\% \pm 2.8\%$ & $8.3\% \pm 2.9\%$  \\
1.2 $<y<$ 2.2 & 0.935 $\pm$ 0.020 & $7.3\% \pm 3.7\%$  & $7.8\% \pm 3.9\%$  \\
\end{tabular} \end{ruledtabular}
\end{table}

\begin{table}[!htb]
\caption{\label{tab:final_results} Fraction of $B$-meson decays in \jpsi 
samples obtained in \pp collisions at \full.}
\begin{ruledtabular} \begin{tabular}{lccc}  
    & \bfrac \\\hline
    -2.2 $<y<$ -1.2 & 8.3\% $\pm$ 2.9\%(stat) $\pm$ 1.9\%(syst)\\
    1.2 $<y<$ 2.2 & 7.8\% $\pm$ 3.9\%(stat) $\pm$ 1.9\%(syst)\\
\\
    1.2 $<|y|<$ 2.2 & 8.1\% $\pm$ 2.3\%(stat) $\pm$ 1.9\%(syst)\\
    \end{tabular} \end{ruledtabular} 
\end{table}

The final results are summarized in Table~\ref{tab:final_results}. Because 
the \pp system is a symmetric, the results from the two arms 
are combined into a statistical average, giving a fraction of \jpsi from 
$B$-meson decays in the 1.2 $<|y|<$ 2.2 region of $8.1\% \pm 2.3\%( 
\rm{stat}) \pm 1.9\%( \rm{syst})$. This result is integrated in the interval
 $0<p_{T}(J/\psi)<$5 GeV/$c$.

\begin{figure}[!htb]
    \includegraphics[width=0.96\linewidth]{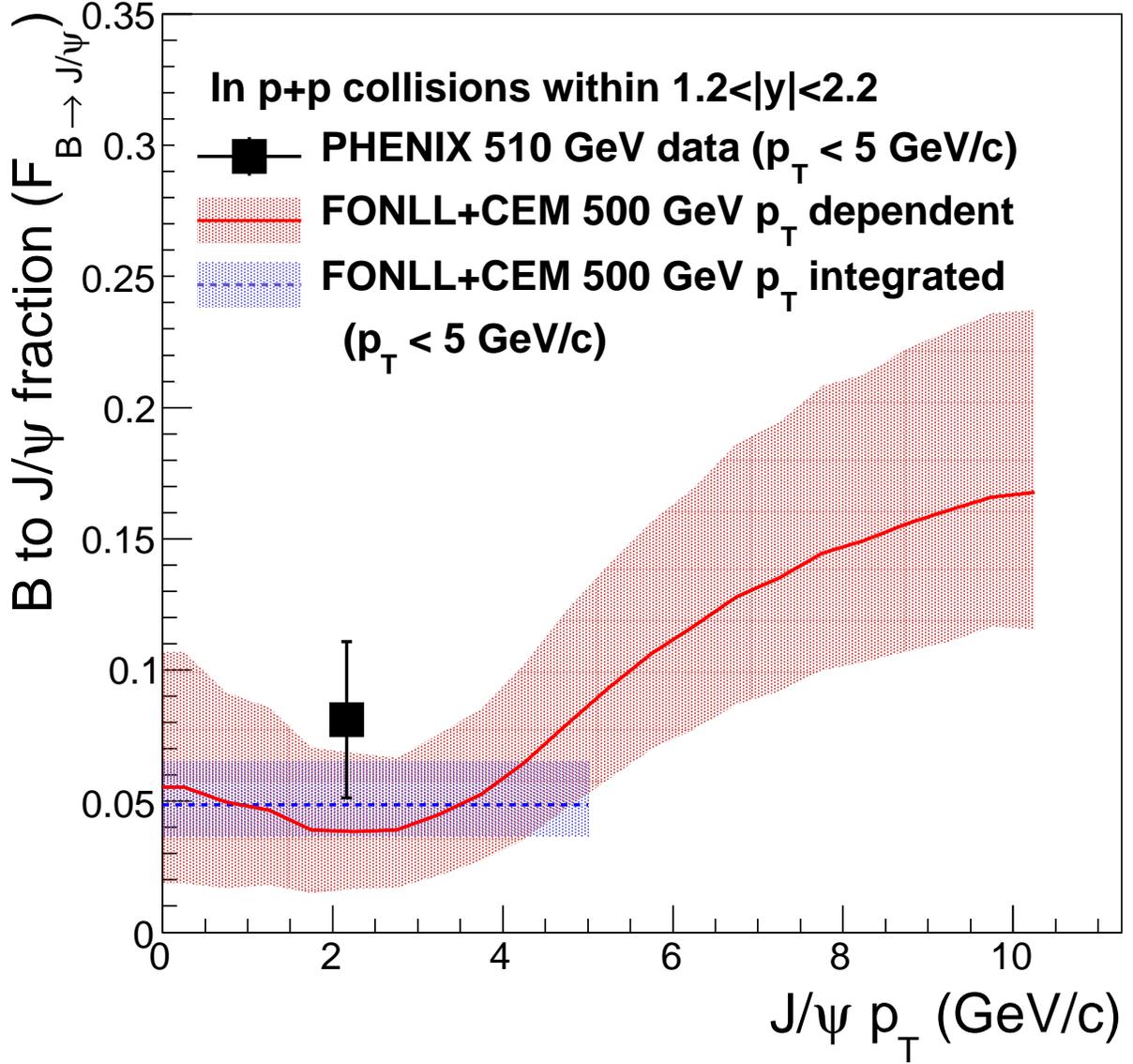}
\caption{\label{fig:fonll_cem} In $p$+$p$ collisions at $\sqrt{s}$ = 510 
GeV and $1.2<|y|<2.2$ rapidity region, comparison of PHENIX $B$ 
$\rightarrow$ $J/\psi$ fraction ($F_{B{\rightarrow}J/\psi}$) measured in integrated \jpsi 
$p_{T}$ range of $p_{T}<5$ \gevc with \jpsi $p_{T}$ dependent (shown 
in solid red) and $p_{T}$ integrated within 0--5 GeV$/c$ region (shown 
in dashed blue) $B$ $\rightarrow$ $J/\psi$ fraction predicted by the 
{\sc fonll+cem}~\cite{Bedjidian2004gd,PhysRevLett.95.122001,vogt_dis} model in 500 GeV 
$p$+$p$ collisions. The uncertainty of the PHENIX measurement is 
statistical and systematic combined.}
\end{figure}

\begin{figure}[!htb]
    \includegraphics[width=0.96\linewidth]{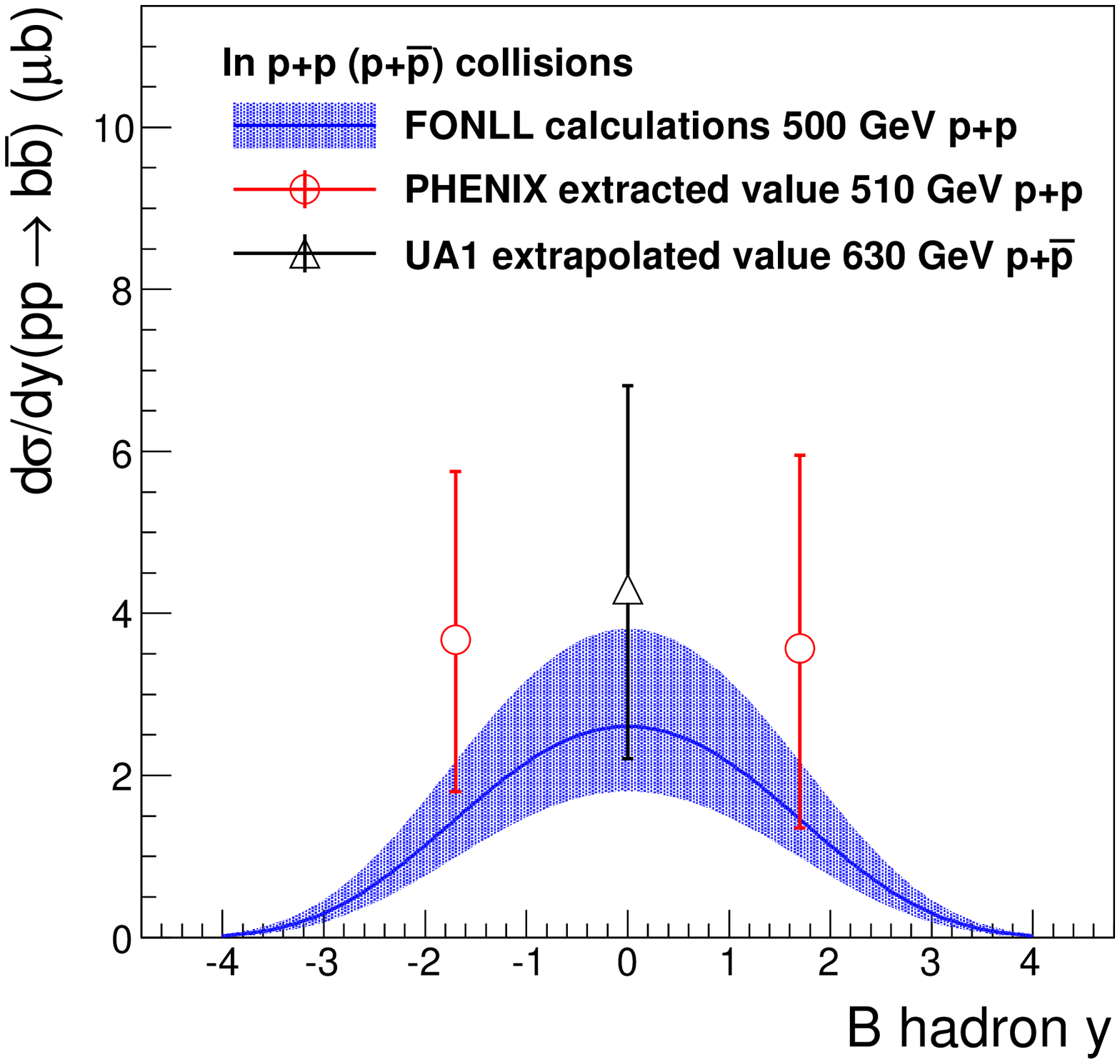}
\caption{\label{fig:fonll_xsec} The average $b\bar{b}$ cross section per 
unit rapidity ($d\sigma/dy(pp{\rightarrow}b\bar{b}+X)$) is determined by the 
$B \rightarrow$ $J/\psi$ fraction (\bfrac) discussed in the paper and the 
inclusive \jpsi cross section in 510 GeV $p$+$p$ collisions extrapolated
from PHENIX 200 GeV $p$+$p$ measurements and the energy scaling factor 
provided by the {\sc cem}~\cite{vogt_dis}. The extrapolated 
$d\sigma/dy(pp{\rightarrow}b\bar{b})$ (shown as open red circles) at $B$ hadron 
mean rapidity $y = \pm1.7$ in 510 GeV $p$+$p$ collisions is compared with 
the rapidity dependent $B$ cross section (shown as blue solid line) 
calculated in {\sc fonll}. The PHENIX result is also comparable with 
the value of UA1 630 GeV $p$+$\bar{p}$ $d\sigma/dy(p\bar{p}{\rightarrow}b\bar{b})$ 
extracted from $p_{T} > 8$ GeV$/c$ to $p_{T} > 0$ range~\cite{ALBAJAR1991121,Albajar1994} 
and unscaled with energy. The uncertainty of the 
extrapolated value at PHENIX (UA1) combines the statistical and 
systematic uncertainty from experiment with the {\sc cem} uncertainty. 
The uncertainty of the {\sc fonll} calculations contains both $b$ quark mass and scaling uncertainties.}
 \end{figure} 

Comparisons to global measurements within the same inclusive \jpsi 
$p_{T}$ region from CDF~\cite{Acosta:2004yw}, ALICE 
\cite{Abelev:2012gx}, CMS~\cite{Khachatryan:2010yr} and LHCb 
\cite{lhcb8TeV, Aaij:2015rla} experiments are shown in 
Fig.~\ref{fig:global_data}(a). The result from PHENIX is also compared with 
the $p_T$-dependent fraction from other experiments using the average 
$p_T=$ 2.2 \gevc of our inclusive \jpsi sample as shown in 
Fig.~\ref{fig:global_data}(b). The LHCb experiment has measurements over a 
wide rapidity range, $2.0<y<4.5$; only results from $2.0<y<2.5$ and 
$3.0<y<3.5$ are shown in Fig.~\ref{fig:global_data}. The $2.0<y<2.5$ 
rapidity range is close to the kinematic range accessed by other 
measurements. The \bfrac result from this measurement is consistent with 
those from the higher energy collisions within uncertainties, although 
it does not exclude the possibility of a decrease of the \bfrac toward 
lower collision energy.

Figure~\ref{fig:fonll_cem} presents the comparison between the 510 GeV 
$p$+$p$ PHENIX result and the fixed-order next-to-leading-log plus 
color-evaporation-model {\sc 
(fonll+cem)}~\cite{Bedjidian2004gd,PhysRevLett.95.122001,vogt_dis} 
predictions for the $B \rightarrow J/\psi$ fraction 
($\textrm{F}_{B{\rightarrow}J/\psi}$) in 500 GeV $p$+$p$ collisions. 
The {\sc cem} $J/\psi$ calculation uses the results of fitting the 
scale parameters to the energy dependence of the open charm total cross 
section for the charm quark mass $m_c = 1.27 \pm 0.09$ GeV$/c^{2}$.  
The factorization and renormalization scales, relative to the mass of 
the charm quark in the total cross section were found to be $\mu_F/m = 
2.1^{+ 2.55}_{-0.85}$ and $\mu_R/m = 
1.6^{+0.11}_{-0.12}$~\cite{vogt_dis}. The same central values were used 
to fix the $J/\psi$ normalization parameter in the {\sc cem} to the 
total cross section at $x_{F}>0$ and $y>0$ as a function of energy.  
The $J/\psi$ distributions were calculated with the same mass and scale 
parameters but to include the $p_{T}$ dependence instead of 
$\mu_{F,R}/m$, $\mu_{F,R}/m_{T}$ was used, where $m_T = 
\sqrt{(p_{T_c}^2 + p_{T_{\overline c}}^2)/2 + m_c^2}$.  The shape of 
the $p_T$ distribution at low $p_T$ is determined by a $k_T$ kick of 
1.29 GeV$/c$ at $\sqrt{s} = 500$ GeV. The energy difference between 500 
GeV and 510 GeV is small, so the difference in the $B \rightarrow 
J/\psi$ fraction is negligible. The measured fraction at PHENIX is 
consistent with the {\sc fonll+cem} model prediction within 
uncertainties. The CMS nonprompt and prompt \jpsi cross section 
measurements at 7 TeV $p$+$p$ collisions~\cite{Khachatryan:2010yr} have 
been compared to the {\sc fonll+cem} calculations as well. The old {\sc 
cem} model underestimated the prompt \jpsi cross section within 
$1.6<|y|<2.4$ and \jpsi $p_{T}<5$ GeV$/c$ region measured by the CMS 
experiment in 7 TeV $p$+$p$ collisions, while the nonprompt \jpsi cross 
section measured in the same kinematic region and experiment is 
consistent with the {\sc fonll} calculations. Calculations with the CEM 
parameters from~\cite{vogt_dis} give a better agreement between the 
{\sc fonll+cem} prediction and the $B \rightarrow J/\psi$ fraction 
measured by CMS~\cite{Khachatryan:2010yr}. The {\sc fonll} calculations 
can reasonably describe the nonprompt \jpsi cross section results at 
LHCb for $p_{T}>0$ ~\cite{lhcb8TeV, Aaij:2015rla}.

The $B \rightarrow J/\psi$ fraction \bfrac is also related to the 
inclusive \jpsi cross section per unit rapidity 
$d\sigma/dy(pp{\rightarrow}J/\psi)$ and the $b\bar{b}$ cross section per unit rapidity 
$d\sigma/dy(pp{\rightarrow}b\bar{b})$, 
\begin{equation}
\label{eq:b_xsec}
F_{B \rightarrow J/\psi} = \frac{2 \times d\sigma/dy(pp{\rightarrow}b\bar{b}) 
\times {\rm{Br}}(B{\rightarrow}J/\psi+X)}{d\sigma/dy(pp{\rightarrow}J/\psi)},
\end{equation}
where ${\rm{Br}}(B{\rightarrow}J/\psi+X)$ is the branching ratio of $B$ 
hadron decays to $J/\psi$ and the $b$ ($\bar{b}$) quark to 
$B$-hadron fragmentation is assumed to be 1. The factor of two in 
Eq. (\ref{eq:b_xsec}) accounts for the fact that both 
$B{\rightarrow}J/\psi$ and $\overline{B}{\rightarrow}J/\psi$ contribute to 
the $B{\rightarrow}J/\psi$ fraction $F_{B{\rightarrow}J/\psi}$.
Eq.(\ref{eq:b_xsec}) can be rewritten as:

\begin{equation}
\label{eq:b_xsec2}
d\sigma/dy(pp{\rightarrow}b\bar{b}) 
= \frac{\frac{1}{2} \times d\sigma/dy(pp{\rightarrow}J/\psi) 
\times F_{B{\rightarrow}J/\psi}}{{\rm{Br}}(B{\rightarrow}J/\psi + X)}.
\end{equation}
Therefore, $d\sigma/dy(pp{\rightarrow}b\bar{b})$ can be derived from 
Eq. (\ref{eq:b_xsec2}). To do this, we use  
$d\sigma/dy(pp{\rightarrow}J/\psi) = 1.00 \pm 0.11$ $\mu$b ($0.97 \pm 0.11$ $\mu$b) 
at mean rapidity $y = 1.7$ ($-1.7$) in 510 GeV $p$+$p$ collisions, and 
${\rm{Br}}(B{\rightarrow}J/\psi + X) = 1.094 \pm 0.032 
\%$~\cite{pdg2016}. Here, $d\sigma/dy(pp{\rightarrow}J/\psi, 510$ GeV$)$ 
is extrapolated as $d\sigma/dy(pp{\rightarrow}J/\psi, 200$ GeV$) \times 
R(510/200)$, where the scaling factor $R(510/200)$ is 
$2.08^{+0.75}_{-0.55}$ according to the {\sc cem}~\cite{vogt_dis}, and 
$d\sigma/dy(pp{\rightarrow}J/\psi, 200$ GeV)$=0.48{\pm}0.05\ {\mu}$b 
($0.47{\pm}0.05\ {\mu}$b) at mean rapidity $y=1.7$ ($-1.7$)~\cite{Adare:2011vq}. 

The extracted $d\sigma/dy(pp{\rightarrow}b\bar{b})$ is 
$3.57^{+2.38}_{-2.22}$ ($3.68^{+2.08}_{-1.88}$) $\mu$b at $B$ hadron 
mean rapidity = 1.7 ($-1.7$) in 510 GeV $p$+$p$ collisions. The 
weighted average of the two measurements is 
$d\sigma/dy(pp{\rightarrow}b\bar{b})=3.63^{+1.92}_{-1.70}$ ${\mu}$b at 
$B$-hadron rapidity$={\pm}1.7$. As shown in Fig.~\ref{fig:fonll_xsec}, 
these values are comparable with the {\sc fonll}-calculated 
rapidity-dependent $B$ cross section within large 
uncertainties~\cite{FONLL,Cacciari:2001td,Cacciari:2012ny}. The PHENIX 
extracted values are also comparable to the UA1 $\sqrt{s}=630$ GeV 
$p$+$\bar{p}$ average $b\bar{b}$ cross section per unit rapidity 
($d\sigma/dy(p\bar{p}{\rightarrow}b\bar{b}, 630$ GeV$) = 
4.3^{+2.51}_{-2.10}$ $\mu$b) within $|y|<1.5$ 
\cite{ALBAJAR1991121,Albajar1994} which is extrapolated from $p_{T}>8$ 
GeV$/c$ to the $p_{T}>0$ range. The {\sc fonll} calculation assumes 
$m_{b} = 4.75 \pm 0.25$ GeV$/c^{2}$ while the renormalization and 
factorization scales are varied by a factor of two around the central 
value, $\mu_{R,F} = 
\sqrt{p_{T}^{2}+m_{b}^{2}}$~\cite{PhysRevLett.95.122001,Cacciari:2012ny}.

\section{Summary}
\label{sec: Summary}

We have presented a new measurement of the nonprompt over inclusive 
\jpsi production ratio \bfrac in \pp collisions at $\sqrt{s}$ = 510 
GeV, integrated over the \jpsi kinematical domain, $p_{T}<5$ GeV/$c$ 
and rapidity $1.2<|y|<2.2$. The result is \bfrac = $8.1\% \pm 2.3\% \  \rm 
(stat) \pm 1.9\% \  \rm (syst)$. This measurement extends the previously 
measured \bfrac values at CDF and LHC to lower energy, and is comparable 
to measurements at higher energies; it is also 
within 1.0 standard deviation of the {\sc fonll+cem} calculation which has 
a nonnegligible dependence on $\sqrt{s}$, $p_{T}$ and $y$. 
The extrapolated $d\sigma/dy(pp{\rightarrow}b\bar{b})$ is 
$3.63^{+1.92}_{-1.70}$ ${\mu}$b at $B$ hadron mean rapidity, 
${\pm}1.7$, in 510 GeV $p$+$p$ collisions,
which is comparable with the 
{\sc fonll} calculations in 500 GeV $p$$+$$p$ collisions.

The weak dependence on the center of mass energy in 
Fig.~\ref{fig:global_data}(a) for the \bfrac fraction could indicate that the 
variation of the bottom yield with energy is compensated by a similar 
variation of the prompt \jpsi yield. It is also noteworthy that only a 
factor of two decrease of the $b$ over the $c$ yield is expected going 
from LHC energies to $\sqrt{s}$ = 510 GeV, as calculated with {\sc fonll} 
~\cite{FONLL,Cacciari:2001td}.  However, modeling the 
hadronization of the bound \cc at low $p_T$ is still a challenge to QCD 
calculations. The present results provide complementary information to 
the surprisingly weak evolution of \bfrac in $0.51 \le \sqrt{s} \le 13$ 
TeV domain, for central or near central rapidity and low $p_{T}$ 
production.

The analysis procedure developed in this study will be applied to other 
data sets recorded by PHENIX at different center of mass energies.  A 
similar method can also be applied to the study of $B$- and $D$-meson 
semileptonic decays to muons, which will help to understand the 
production mechanism of charm and bottom, and provide a complementary 
measurement to the one presented in this paper.


\section*{ACKNOWLEDGMENTS}


We thank the staff of the Collider-Accelerator and Physics
Departments at Brookhaven National Laboratory and the staff of
the other PHENIX participating institutions for their vital
contributions.  We acknowledge support from the 
Office of Nuclear Physics in the
Office of Science of the Department of Energy,
the National Science Foundation, 
Abilene Christian University Research Council, 
Research Foundation of SUNY, and
Dean of the College of Arts and Sciences, Vanderbilt University 
(U.S.A),
Ministry of Education, Culture, Sports, Science, and Technology
and the Japan Society for the Promotion of Science (Japan),
Conselho Nacional de Desenvolvimento Cient\'{\i}fico e
Tecnol{\'o}gico and Funda\c c{\~a}o de Amparo {\`a} Pesquisa do
Estado de S{\~a}o Paulo (Brazil),
Natural Science Foundation of China (People's Republic of China),
Croatian Science Foundation and
Ministry of Science, Education, and Sports (Croatia),
Ministry of Education, Youth and Sports (Czech Republic),
Centre National de la Recherche Scientifique, Commissariat
{\`a} l'{\'E}nergie Atomique, and Institut National de Physique
Nucl{\'e}aire et de Physique des Particules (France),
Bundesministerium f\"ur Bildung und Forschung, Deutscher
Akademischer Austausch Dienst, and Alexander von Humboldt Stiftung (Germany),
National Science Fund, OTKA, K\'aroly R\'obert University College, 
and the Ch. Simonyi Fund (Hungary),
Department of Atomic Energy and Department of Science and Technology (India), 
Israel Science Foundation (Israel), 
Basic Science Research Program through NRF of the Ministry of Education (Korea),
Physics Department, Lahore University of Management Sciences (Pakistan),
Ministry of Education and Science, Russian Academy of Sciences,
Federal Agency of Atomic Energy (Russia),
VR and Wallenberg Foundation (Sweden), 
the U.S. Civilian Research and Development Foundation for the
Independent States of the Former Soviet Union, 
the Hungarian American Enterprise Scholarship Fund,
and the US-Israel Binational Science Foundation.


%
 
\end{document}